\documentclass[prb,twocolumn,amssymb,showpacs,showkeys,floatfix,a4paper]{revtex4}
\usepackage[dvips]{graphicx}
\usepackage{times}
\usepackage[a4paper=true,pagebackref=true]{hyperref}
\hypersetup{
  pdftitle = XAFS study on GaFeN(:Si),
  pdfauthor = Mauro Rovezzi,
  pdfsubject= diluted magnetic semiconductors,
  pdfkeywords = GaFeN EXAFS DMS,
  colorlinks = true,
  linkcolor = blue,
  anchorcolor = blue,
  citecolor = blue,
  filecolor = blue,
  pagecolor = blue,
  urlcolor = blue
}

\begin{document}

\title{Local structure of (Ga,Fe)N and (Ga,Fe)N:Si investigated by x-ray absorption fine structure spectroscopy}

\author{Mauro Rovezzi}\email{mauro.rovezzi@esrf.eu}
\affiliation{Italian Collaborating Research Group, BM08 ``GILDA'' - ESRF, BP 220, F-38043 Grenoble, France}
\affiliation{Consiglio Nazionale delle Ricerche - Operative Group in Grenoble (INFM-OGG), BP 220, F-38043 Grenoble, France}

\author{Francesco D'Acapito}
\affiliation{Italian Collaborating Research Group, BM08 ``GILDA'' - ESRF, BP 220, F-38043 Grenoble, France}
\affiliation{Consiglio Nazionale delle Ricerche - Operative Group in Grenoble (INFM-OGG), BP 220, F-38043 Grenoble, France}

\author{Andrea Navarro-Quezada}
\affiliation{Institute for Semiconductor and Solid State Physics, Johannes Kepler University, A-4040, Linz, Austria}

\author{Bogdan Faina}
\affiliation{Institute for Semiconductor and Solid State Physics, Johannes Kepler University, A-4040, Linz, Austria}

\author{Tian Li}
\affiliation{Institute for Semiconductor and Solid State Physics, Johannes Kepler University, A-4040, Linz, Austria}

\author{Alberta Bonanni}
\affiliation{Institute for Semiconductor and Solid State Physics, Johannes Kepler University, A-4040, Linz, Austria}

\author{Francesco Filippone}
\affiliation{Istituto di Struttura della Materia (ISM), Consiglio Nazionale delle Ricerche, Via Salaria Km 29.5, CP
10, 00016 Monterotondo Stazione, Italy}

\author{Aldo Amore Bonapasta}
\affiliation{Istituto di Struttura della Materia (ISM), Consiglio Nazionale delle Ricerche, Via Salaria Km 29.5, CP
10, 00016 Monterotondo Stazione, Italy}

\author{Tomasz Dietl}
\affiliation{Institute of Physics, Polish Academy of Sciences, PL-02-668 Warszawa, Poland}
\affiliation{Institute of Theoretical Physics, University of Warsaw, PL-00-681 Warszawa, Poland}

\date{\today}

\begin{abstract}
 X-ray absorption fine-structure (XAFS) measurements supported by {\em ab initio} computations within the density functional theory (DFT) are employed to systematically characterize Fe-doped as well as Fe and Si-co-doped  films grown by metalorganic vapour phase epitaxy. The analysis of extended-XAFS data shows that depending on the growth conditions, Fe atoms either occupy Ga substitutional sites in GaN or precipitate in the form of $\epsilon$-Fe$_3$N nanocrystals, which are ferromagnetic and metallic according to the DFT results. Precipitation can be hampered by reducing the Fe content, or by increasing the growth rate or by co-doping with Si. The near-edge region of the XAFS spectra provides information on the Fe charge state and shows its partial reduction from Fe$^{+3}$ to Fe$^{+2}$ upon Si co-doping, in agreement with the Fe electronic configurations expected within various implementations of DFT.
\end{abstract}



\maketitle

\section{Introduction}

 It has been realized in the recent years that a palette of nanocharacterization tools has to be employed in order to elucidate the origin of the surprising high-temperature ferromagnetism detected in a number of magnetically doped semiconductors and oxides.\cite{Kuroda:2007_NM,Bonanni:2007_SST,Dietl:2007_JPCM,Chambers:2006_MT,Ando:2008_PRL} The application of state-of-the art analytic methods appears to reveal that the distribution of magnetic cations is generally highly non-uniform and that there is an ultimate relationship between the ion arrangement and the magnetic response of the system. In particular, randomly distributed localized spins account for the paramagnetic component of the magnetization, whereas regions with a high local density of magnetic cations are presumably responsible for ferromagnetic features. It has been found that the aggregation of magnetic ions -- driven by a substantial contribution of open $d$ shells to the bonding -- leads either to crystallographic phase separation, {\i.e.}, to the precipitation of a magnetic compound or an elemental ferromagnet, or to the chemical phase separation into alternating regions with higher and lower concentration of magnetic cations, occurring without distorting the crystallographic structure. In the literature on diluted magnetic semiconductors (DMS), this chemical phase separation is commonly referred to as spinodal decomposition,\cite{Kuroda:2007_NM,Katayama-Yoshida:2007_PSSA} independently of its microscopic origin.

 There are already experimental indications that the above scenario applies to the case of (Ga,Fe)N fabricated by means of metalorganic vapor phase epitaxy (MOVPE),\cite{Bonanni:2007_PRB,Pacuski:2008_PRL,Bonanni:2008_PRL} where -- depending on the growth parameters and on the Fe concentration -- Fe is either randomly distributed over cation sites or non-uniformly incorporated, showing spinodal decomposition or the presence of precipitates that have been identified as hexagonal $\epsilon-$Fe$_3$N, according to synchrotron radiation x-ray diffraction (XRD) and high-resolution transmission electron microscopy (HRTEM).\cite{Bonanni:2007_PRB,Bonanni:2008_PRL} Moreover, by employing these nanocale characterization methods it has been found that the aggregation of Fe cations is largely hampered by co-doping with Si donors or Mg acceptors.\cite{Bonanni:2008_PRL}

 In this work, we present results of x-ray absorption fine structure (XAFS) spectroscopy\cite{Lee:1981_RMP} carried out on (Ga,Fe)N and (Ga,Fe)N:Si samples. XAFS is a well established tool in the study of semiconductor heterostructures and nanostructures \cite{Boscherini:2008_book} and has proven its power as a chemically sensitive local probe for the site identification of Mn/Fe dopants in III-V DMS,\cite{Soo:2001_JSR,Soo:2001_APL,Soo:2004_APL,Bacewicz:2005_JPCS,D'Acapito:2006_PRB} and of Mn-rich nano-columns in Ge.\cite{Rovezzi:2008_APL} In the present case it allows us to obtain information on the local atom arrangement in the  vicinity of Fe ions as well as on the Fe charge state, depending on the Fe concentration, growth conditions, and co-doping. We then compare the experimentally determined bond lengths to the corresponding values expected from {\em ab initio} simulations. This combined experimental and computational effort makes it possible to identify the epitaxial parameters controlling the formation of $\epsilon$-Fe$_3$N precipitates in (Ga,Fe)N. However, no presence of bonds specific to $\epsilon$-Fe$_3$N has been detected in the (Ga,Fe)N films co-doped with Si. This finding provides a strong corroboration of the previous theoretical suggestions\cite{Dietl:2006_NM} and experimental observations\cite{Bonanni:2008_PRL} indicating that the alteration of the magnetic ions valence by shallow impurities can hinder their aggregation and, therefore, extend the solubility range.

 Our paper is organized as follows. In Sec.~\ref{sec:samples} we present the relevant characteristics of the studied samples and we briefly recall the employed growth and doping procedures. The XAFS method and the analysis of both the Extended (EXAFS) and Near-Edge (XANES) regions of collected spectra are described in Sec.~\ref{sec:xafs}. The employed {\em ab initio} computation scheme are summarized in Sec.~\ref{sec:theo} for the case of wurtzite (Ga,Fe)N and $\epsilon$-Fe$_3$N. Section~\ref{sec:results} contains an interpretation of the experimentally determined values of the bond lengths in the light of the theoretical results. As discussed in Sec.~\ref{sec:discussion}, this insight allows to quantify the relative concentration of Fe in the $\epsilon$-Fe$_3$N precipitates depending on the growth conditions and co-doping with Si. The main conclusions and outlook of our work are summarized in Sec.~\ref{sec:conclusions}.

\section{Experimental}

\subsection{Samples production}\label{sec:samples}

 The (Ga,Fe)N and (Ga,Fe)N:Si epilayers considered here are fabricated by MOVPE on $c$-plane sapphire substrates following the growth procedures and applying the {\em in situ} and {\em ex situ} characterization methods we have formerly reported for (Ga,Fe)N.\cite{Bonanni:2007_PRB} The total Fe concentration in the samples ranges from 4$\times$10$^{19}$~cm$^{-3}$ to 3$\times$10$^{20}$~cm$^{-3}$ for a Fe-precursor (Cp$_{2}$Fe) flow-rate from 50 to 350 standard cubic centimeters per minute (sccm).\cite{Bonanni:2007_PRB} In the Si-doped (Ga,Fe)N structures the Si-content is estimated to be 1$\times$10$^{19}$~cm$^{-3}$. The growth-rate during deposition is regulated by the Ga-precursor (TMGa) flow-rate and varies from 0.08~nm/s for 5~sccm to 0.3~nm/s for 12~sccm of TMGa flow.

\subsection{Experiments and data analysis}\label{sec:xafs}

 The XAFS measurements at the Fe-K edge (7112~eV) are carried out at the ``GILDA'' Italian collaborating research group beamline (BM08) at the European Synchrotron Radiation Facility in Grenoble.\cite{D'Acapito:1998_EN} The monochromator is equipped with a pair of Si(111) crystals and run in dynamical focusing mode.\cite{Pascarelli:1996_JSR} Harmonics rejection is achieved by using a pair of Pd-coated mirrors with an estimated cutoff of 18~keV. Data are collected in the fluorescence mode (normal geometry) using a 13-element hyper pure Ge detector and normalized by measuring the incident beam with an ion chamber filled with nitrogen gas. In order to minimize the effects of coherent scattering from the substrate, the samples are mounted on a vibrating sample holder\cite{Tullio:2001_tech} and measurements are carried out at liquid nitrogen temperature to reduce thermal disorder. For each sample the integration time for each energy point and the number of acquired spectra are chosen in order to collect $\approx 10^6$ counts on the final averaged spectrum. In addition, before and after each measurement a metallic Fe reference foil is measured in transmission mode to check the stability of the energy scale and to provide an accurate calibration. In this way we locate at 7112.0~eV the first inflection point of the absorption spectrum.\cite{Bearden:1967_RMP}

 The EXAFS signal - $\chi(k)$ - is extracted from the absorption raw data - $\mu(E)$ - with the {\sc viper} program\cite{Klementev:2001_JPDAP} employing a smoothing spline algorithm and choosing the energy edge value ($E_0$) at the maximum derivative. The quantitative analysis is carried out with the {\sc i\-feffit-artemis} programs\cite{Newville:2001_JSR,Ravel:2005_JSR} with the atomic models described below. Theoretical EXAFS signals are computed with the {\sc feff8.4} code\cite{Ankudinov:1998_PRB} using muffin tin potentials and the Hedin-Lunqvist approximation for their energy-dependent part. The free fitting parameters used in the analysis are: $S_0^2$ (common amplitude parameter), $\Delta E_0$ (the refinement of the edge position), $R_i$ and $\sigma^2_i$, the interatomic distance and the Debye-Waller factor for the $i^{th}$ atomic shell around Fe respectively. In addition, a linear combination parameter ($X$) is fitted when two theoretical models are used in the same fit. The fits are carried out in the Fourier-transformed space (R space) in the range [1.3 -- 3.5]~\AA\ from the $k^2$-weighted EXAFS data in the range [2.5 -- 10.5]~\AA$^{-1}$ using Hanning windows with slope parameter $dk = 1$ and $dR = 0.1$, for the forward and backward Fourier transforms, respectively. In order to reduce the variables used in the fit and keep the theoretical models coherent for all samples, the fits are limited to $R = 3.5$~\AA. In fact, the substitutional site in the wurtzite lattice can be reduced to two average distances of Fe--N and Fe--Ga and for $\epsilon$-Fe$_3$N to the two average distances Fe--N and Fe--Fe. In addition, for simplicity, only one average Fe--N distance is reported in the text. On the other hand, for samples that present only the substitutional phase it is possible to expand the model to upper distances including multiple scattering (MS) contributions (as pointed out, for example, in the {\em Wurtzite phase analysis} section of Ref.~\onlinecite{Decremps:2003_PRB}) but the results are found to be equivalent. In the case of mixed phases, the expansion to longer distances and MS paths, reduces considerably the overall quality of the fits due to the lack of a model compound for $\epsilon$-Fe$_3$N.

 The XANES spectra are normalized using the {\sc athena} program,\cite{Ravel:2005_JSR} setting the edge jump value to unity. The peaks appearing in the energy region before the edge are analyzed with the {\sc fityk} program\cite{Wojdyr:_misc} to a curve consisting in an arctangent background\cite{Note:atan} plus one or two pseudo-Voigt peaks.\cite{Galoisy:2001_CG,Westre:1997_JACS}

\subsection{Theoretical methods}\label{sec:theo}

 The properties of a Fe atom substituting a Ga atom in GaN (Fe$_{\mathrm{Ga}}$) as well as of the $\epsilon$-Fe$_3$N hexagonal phase are investigated by density functional theory (DFT) methods by using both the local spin density - generalized gradient approximation (LSD-GGA)\cite{Perdew:1996_PRL} and the LSD-GGA+U formalism\cite{Anisimov:1997_JPCM} as implemented with plane wave basis sets in {\sc quantum-espresso},\cite{Cococcioni:2005_PRB,Kulik:2006_PRL,Giannozzi:_misc} in order to take into account the strong localization of the $d$ states of Fe, poorly described by LSD-GGA exchange-correlation functionals. Total energies are calculated  in a supercell approach, by using ultrasoft pseudopotentials\cite{Vanderbilt:1990_PRB} planewave basis sets, the special-points technique for {\bf k}-space integration, and the Perdew-Burke-Ernzerhof (PBE) exchange-correlation functional.\cite{Perdew:1996_PRL}

 In detail, for Fe$_{\mathrm{Ga}}$ the (1,1,1) {\bf k}-point Monkhorst-Pack mesh for a 72-atoms wurtzite supercell of GaN (corresponding to 3x3x2 unit cells), a gaussian smearing of the occupation numbers and plane waves cutoffs of 25 Ry for wavefunctions and 150 Ry for densities are used. The electronic channels considered in the atomic pseudopotentials are $2s$ and $2p$ for N, $3s$, $3p$, $3d$, $4s$, $4p$ for Fe, and $3d$, $4s$, $4p$ for Ga. One substitutional Fe atom is included in the 72-atoms supercell of GaN. A neutralizing background charge is imposed when dealing with charged states of Fe$_{\mathrm{Ga}}$. Geometry optimizations are performed by fully relaxing the positions of all the supercell atoms by minimizing the atomic forces. The spin state of the system is self-consistently determined during the wavefunction optimization. The position of electronic levels induced by the Fe impurity in the GaN energy gap is estimated by calculating the corresponding transition energy levels $\epsilon^{n/n+1}$, that is, the Fermi-energy values for which the charge of the defect changes from $n$ to $n+1$; this gives the position of the Fermi energy (E$_{\mathrm{F}}$) with respect to the top of the valence band. These values are estimated from total energies (defect formation energies) as described in Refs.~\onlinecite{Bonapasta:2003_PRB,Walle:2004_JAP}, where further details on the theoretical methods can be found. Transition energy values have to be located with respect to the GaN energy gap estimated here by the $\epsilon^{0/-}$ transition level relative to bulk GaN, thus permitting to compare defect transition levels with an energy gap calculated in a consistent way.\cite{Amore:2004_PRB} Kohn-Sham electronic eigenvalues at the $\Gamma$ point and electronic density of states (DOS) have been also considered when discussing the electronic properties of the  Fe$_{\mathrm{Ga}}$ impurity.

 In the case of $\epsilon$-Fe$_3$N, the starting point of the calculations is the hexagonal phase as found in literature on the nitridation process of Fe.\cite{Jack:1952_AC,Jacobs:1995_JAC} The space group is P6322 (No.~182) with the unit cell composed of 6 Fe atoms in the Wickoff site 6g [($x$, 0, 0); (0, $x$, 0); (-$x$, -$x$, 0); (-$x$, 0, 1/2); (0, -$x$, 1/2); ($x$, $x$, 1/2)] where $x$ = 0.333 and 2 N in site 2c [(1/3, 2/3, 1/4); (2/3, 1/3, 3/4)], that is, iron atoms show the motif of a slightly distorted hexagonal close packing (hcp) structure and nitrogen atoms occupy only corner-sharing octahedra (Fig.~7 in Ref.~\onlinecite{Jacobs:1995_JAC}). Satisfactorily convergent results are achieved by using the (12,12,8) {\bf k}-point Monkhorst-Pack mesh, plane waves cutoffs of 35~Ry for wavefunctions and 140~Ry for densities.

\section{Results}\label{sec:results}

\begin{table*}
 \caption{Results of the EXAFS quantitative analysis for the first (Ga,Fe)N series at TMGa = 5 sccm (A), the second (Ga,Fe)N series at Cp$_2$Fe = 300 sccm (B) and  the third (Ga,Fe)N:Si series as a function of both TMGa and Cp$_2$Fe (C); for clarity only average distances are reported (see text). The values of $S_0^2$ and $\Delta E_0$ are stable, in the limit of error bars, at 0.9(1) and 6(3)~eV, respectively. Averaged crystallographic distances are also reported from literature (D) for comparison. The last part of the table (E), report LSD+U atomic distances calculated as described in the text. Errors on the last significative digit are given in brackets.\cite{Note:errors}}\label{tab:dists}
\begin{tabular}{|cc|cc|cc|cc|c|}
\hline
\hline
Cp$_2$Fe & TMGa & \multicolumn{2}{|c|}{$<$Fe--N$>$} & \multicolumn{2}{|c|}{$<$Fe--Fe$>$} & \multicolumn{2}{|c|}{$<$Fe--Ga$>$} & $X$\\
\multicolumn{2}{|c|}{} & $R_1$ & $\sigma_1$ & $R_2$ & $\sigma_2$ & $R_3$ & $\sigma_3$ & \\
(sccm) & (sccm) & (\AA) & ($10^{-3}$\AA $^{-2}$) & (\AA) & ($10^{-3}$\AA $^{-2}$) & (\AA) & ($10^{-3}$\AA $^{-2}$) & (\%)\\
\hline
\multicolumn{9}{|l|}{A. (Ga,Fe)N as a function of Fe flow-rate}\\
\hline
150 & 5  & 2.02(1) & 6(2)  & ---     & ---   & 3.18(1) & 6(2) & 100(10)\\
200 & 5  & 1.98(1) & 7(1)  & 2.71(3) & 19(5) & 3.19(1) & 7(1) & 70(10) \\
250 & 5  & 1.98(1) & 7(1)  & 2.70(3) & 22(5) & 3.19(1) & 7(1) & 70(10) \\
300 & 5  & 2.01(2) & 8(1)  & 2.74(3) & 21(2) & 3.21(2) & 8(1) & 50(10) \\
350 & 5  & 2.02(3) & 7(2)  & 2.75(4) & 21(2) & 3.22(2) & 7(2) & 40(10) \\
\hline
\multicolumn{9}{|l|}{B. (Ga,Fe)N as a function of Ga flow-rate}\\
\hline
300 & 5  & 2.01(1) & 9(1)  & 2.75(2) & 28(2) & 3.23(1) & 9(1) & 50(10) \\
300 & 8  & 1.98(2) & 12(4) & ---     & ---   & 3.19(1) & 9(1) & 100(10)\\
300 & 10 & 1.97(1) & 8(3)  & ---     & ---   & 3.19(1) & 7(1) & 100(10)\\
300 & 12 & 1.97(1) & 13(5) & ---     & ---   & 3.19(1) & 8(1) & 100(10)\\
\hline
\multicolumn{9}{|l|}{C. (Ga,Fe)N:Si}\\
\hline
300 & 5  & 1.99(2) & 13(3) & ---     & ---   & 3.19(1) & 9(1) & 100(10)\\
300 & 10 & 1.98(2) & 9(2)  & ---     & ---   & 3.18(1) & 7(1) & 100(10)\\
250 & 10 & 1.99(2) & 14(3) & ---     & ---   & 3.18(1) & 8(1) & 100(10)\\
100 & 10 & 2.01(3) & 11(4) & ---     & ---   & 3.19(1) & 9(1) & 100(10)\\
\hline
\multicolumn{9}{|l|}{D. Crystallographic}\\
\hline
\multicolumn{2}{|c|}{$\epsilon$-Fe$_3$N \cite{Jacobs:1995_JAC}} & \multicolumn{2}{|c|}{1.927(1)} & \multicolumn{2}{|c|}{2.703(2)} & \multicolumn{2}{|c|}{---} & ---\\
\multicolumn{2}{|c|}{$\alpha$-Fe \cite{Swanson:1955_NBoS}}       & \multicolumn{2}{|c|}{---} & \multicolumn{2}{|c|}{2.499/2.886} & \multicolumn{2}{|c|}{---} & ---\\
\multicolumn{2}{|c|}{GaN \cite{Schulz:1977_SSC}}                & \multicolumn{2}{|c|}{1.95} & \multicolumn{2}{|c|}{---} & \multicolumn{2}{|c|}{3.18} & ---\\
\hline
\multicolumn{9}{|l|}{E. DFT}\\
\hline
\multicolumn{2}{|c|}{$\epsilon$-Fe$_3$N} & \multicolumn{2}{|c|}{1.89(1)} & \multicolumn{2}{|c|}{2.67(1)} & \multicolumn{2}{|c|}{---}     & ---\\
\multicolumn{2}{|c|}{Fe$_{Ga}^{+1}$}     & \multicolumn{2}{|c|}{1.97(1)} & \multicolumn{2}{|c|}{---}     & \multicolumn{2}{|c|}{3.23(1)} & ---\\
\multicolumn{2}{|c|}{Fe$_{Ga}^{0}$}      & \multicolumn{2}{|c|}{1.99(1)} & \multicolumn{2}{|c|}{---}     & \multicolumn{2}{|c|}{3.22(1)} & ---\\
\multicolumn{2}{|c|}{Fe$_{Ga}^{-1}$}     & \multicolumn{2}{|c|}{2.05(1)} & \multicolumn{2}{|c|}{---}     & \multicolumn{2}{|c|}{3.21(1)} & ---\\
\hline
\hline
\end{tabular}
\end{table*}

\subsection{Solubility limit as a function of Fe content}\label{sec:subres1}

\begin{figure}[!h]
\includegraphics[width=0.49\textwidth]{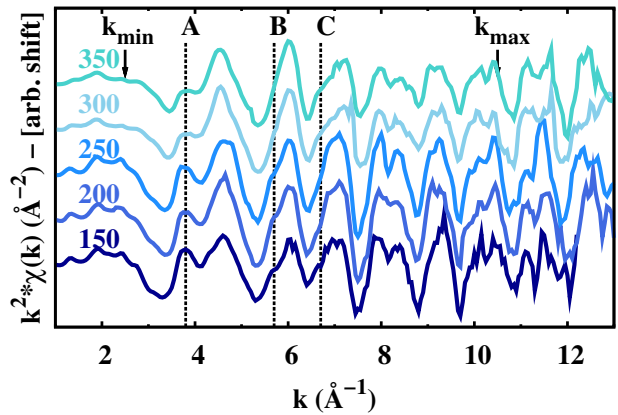}
 \caption{(Color online) $k^2$-weighted EXAFS signals for (Ga,Fe)N samples at TMGa = 5 sccm as a function of Cp$_2$Fe content (sccm, labels on spectra); vertical dashed lines highlight some parts of the spectra as described in Sec.~\ref{sec:subres1}; $k_{min}$ and $k_{max}$ delimit the Fourier-transformed part of the spectrum used in the fit.}\label{fig:gafen1-chi}
\end{figure}

\begin{figure}[!h]
\includegraphics[width=0.49\textwidth]{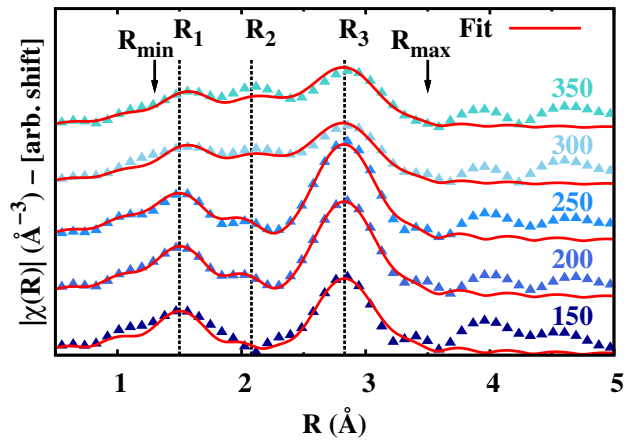}
 \caption{(Color online) Amplitude of the FT for spectra shown in Fig.~\ref{fig:gafen1-chi} with relative fits . $R_{min}$ and $R_{max}$ delimit the fit region; $R_1$, $R_2$ and $R_3$ (with relative dashed vertical lines) indicate the main average distances found in the fit. The R scale has no phase correction, so that all the peaks appear shifted by $\approx 0.4$~\AA.}\label{fig:gafen1-r}
\end{figure}

 The first series systematically studied consists of samples prepared each with different Fe content (Cp$_2$Fe from 150 to 350 sccm) at a fixed growth-rate (TMGa = 5 sccm). The recent investigation by HRTEM \cite{Li:2008_JCG} shows on this series a solubility limit at 200 sccm at our growth conditions. This means that after this limit precipitates appear mostly as $\epsilon$-Fe$_3$N and only at the surface of selected samples as $\alpha$-Fe. In Fig.~\ref{fig:gafen1-chi} are plotted the $k^2$-weighted EXAFS data where qualitative differences are clearly visible and are highlighted by the peaks A, B and C. In particular, three different spectra are visible: for 150 sccm, for 200 and 250 sccm, for 300 and 350 scmm. This difference is also found in the Fourier transformed (FT) spectra shown in Fig.~\ref{fig:gafen1-r} where, in the range $R_{min}$--$R_{max}$, from a two-peaks situation ($R_1$ and $R_3$) at 150 sccm, an intermediate third peak ($R_2$) appears and increases in amplitude with increasing Fe content.

 In the quantitative fit, the data at 150 sccm can be reproduced with a two shell model consisting in a Fe--N ($R_1$) and Fe--Ga ($R_3$) shells. The intermediate peak ($R_2$) is obtained with a Fe--Fe shell. The numerical results are shown in Table~\ref{tab:dists}A.

\subsection{Fe incorporation dependence on growth rate}

\begin{figure}
\includegraphics[width=0.49\textwidth]{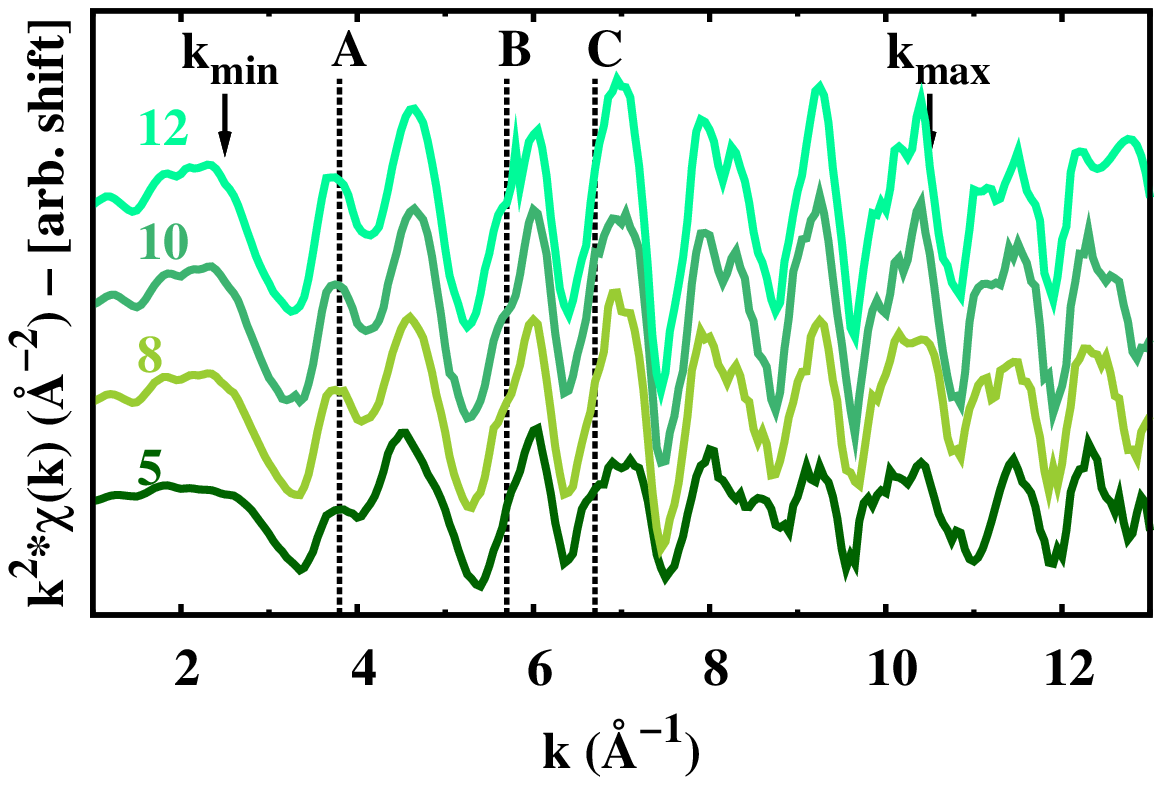}
 \caption{(Color online) $k^2$-weighted EXAFS signals for (Ga,Fe)N samples at Cp$_2$Fe = 300 sccm as a function of TMGa flow-rate (sccm, labels on spectra); vertical dashed lines highlight some parts of the spectra as in Fig~\ref{fig:gafen1-chi}; $k_{min}$ and $k_{max}$ delimit the Fourier-tranformed part of the spectrum used in the fit.}\label{fig:gafen2-chi}
\end{figure}

\begin{figure}
\includegraphics[width=0.49\textwidth]{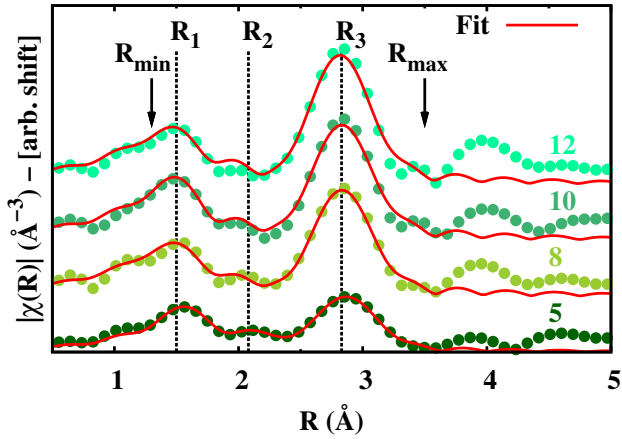}
 \caption{(Color online) Amplitude of the FT for spectra shown in Fig.~\ref{fig:gafen2-chi} with relative fits . $R_{min}$ and $R_{max}$ delimit the fit region; $R_1$, $R_2$ and $R_3$ (with relative dashed vertical lines) indicate the main average distances found from the fit. The R scale has no phase correction, so that all the peaks appear shifted by $\approx 0.4$~\AA.}\label{fig:gafen2-r}
\end{figure}

 The second series studied permits to investigate how the Fe incorporation changes with the growth rate. In fact, by fixing the Fe content at 300 sccm ({\i.e.}, well above the solubility limit estimated to be about 200 sccm), the TMGa flow-rate is increased from 5 to 12 sccm through the samples series. The EXAFS spectra for this series are shown in Fig.~\ref{fig:gafen2-chi} and their respective FTs in Fig.~\ref{fig:gafen2-r}. From a qualitative point of view we note that for high growth rate values (TMGa $\geq 8$ sccm) the EXAFS signal consists in two main frequencies (relative to the distances $R_1$ and $R_3$ in the FT) whereas in samples grown below that limit a further phase is detected and revealed by the peak $R_2$. Also in this case the quantitative analysis is conducted by reproducing data either with a substitutional model or with a combination of substitutional plus $\epsilon$-Fe$_3$N phase. The results are reported in Table~\ref{tab:dists}B.

\subsection{Si co-doping}

\begin{figure}
\includegraphics[width=0.49\textwidth]{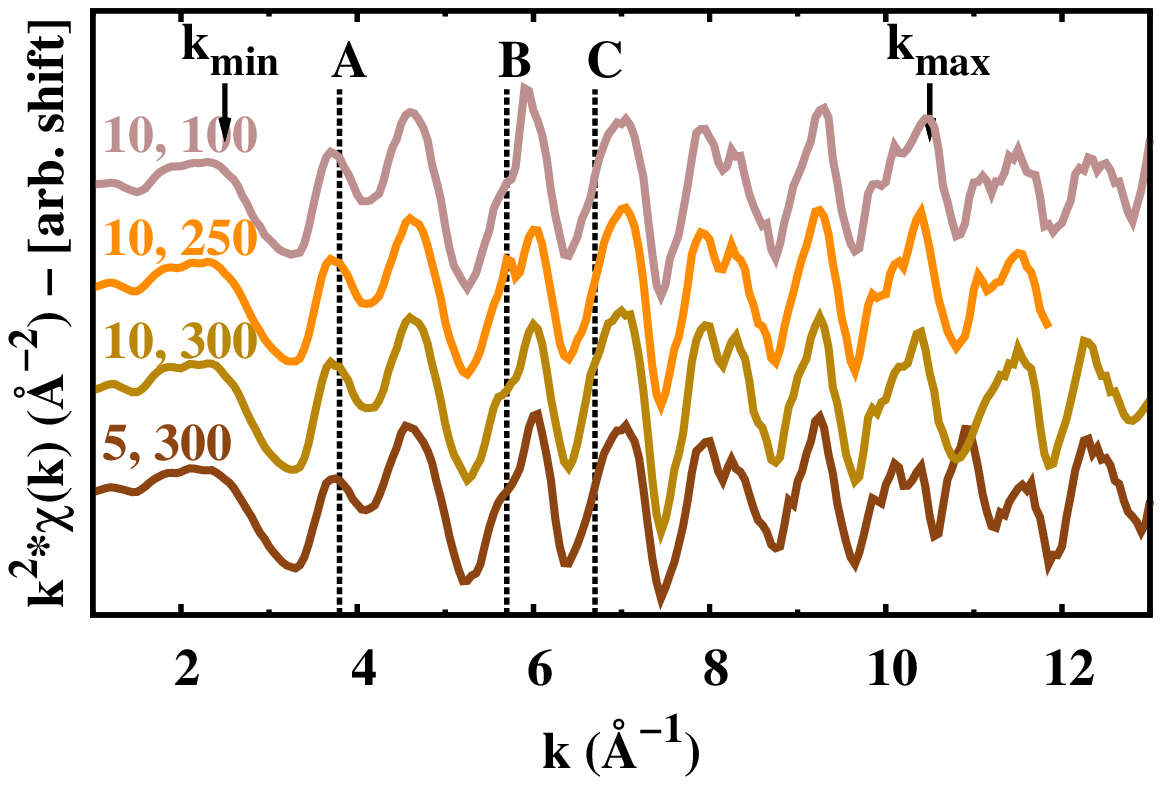}
 \caption{(Color online) $k^2$-weighted EXAFS signals for (Ga,Fe)N:Si samples as a function of TMGa and Cp$_2$Fe (sccm, labels on spectra); vertical dashed lines highlight some parts of the spectra as in Figs.~\ref{fig:gafen1-chi},\ref{fig:gafen2-chi}; $k_{min}$ and $k_{max}$ delimit the Fourier-tranformed part of the spectrum used in the fit.}\label{fig:gafensi-chi}
\end{figure}

\begin{figure}
\includegraphics[width=0.49\textwidth]{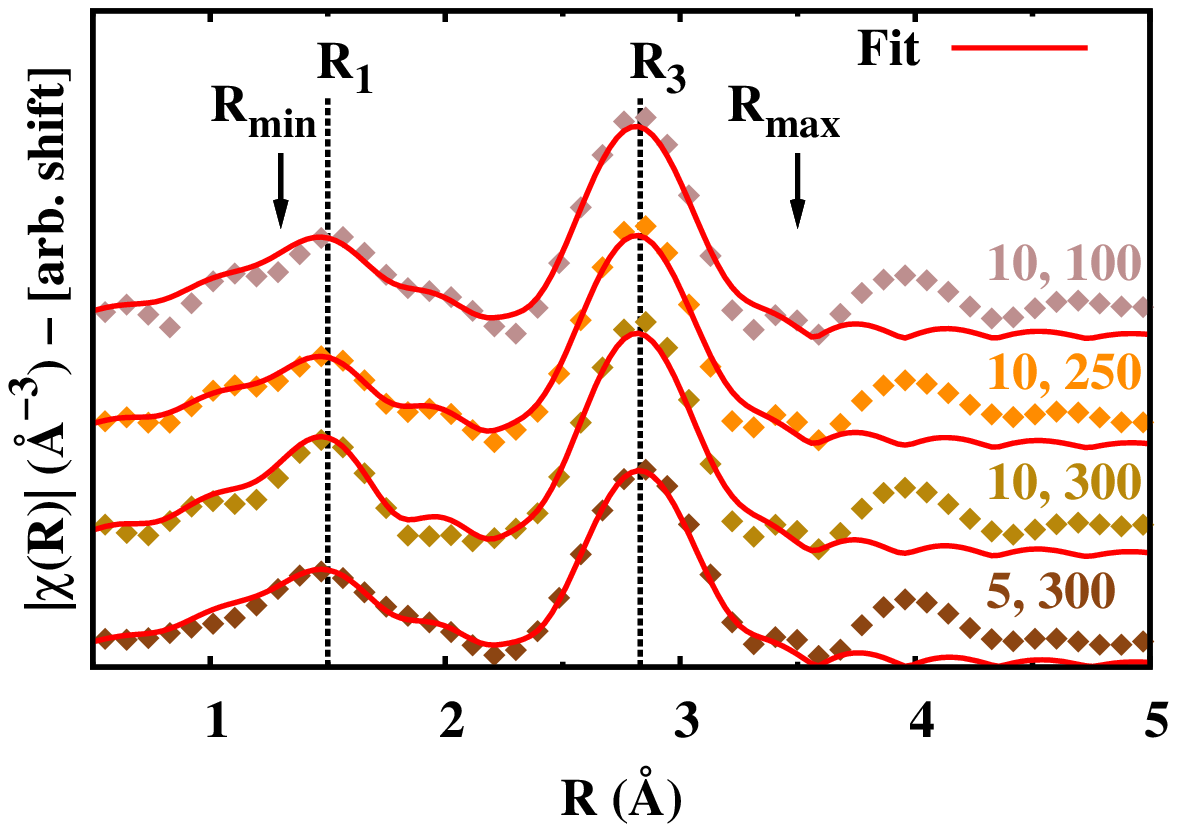}
 \caption{(Color online) Amplitude of the FT for spectra shown in Fig.~\ref{fig:gafensi-chi} with relative fits . $R_{min}$ and $R_{max}$ delimit the fit region; $R_1$ and $R_3$ (with relative dashed vertical lines) indicate the main average distances as in Figs.~\ref{fig:gafen1-r} and \ref{fig:gafen2-r}. The R scale has no phase correction, so that all the peaks appear shifted by $\approx 0.4$~\AA.}\label{fig:gafensi-r}
\end{figure}

 Finally, the effect of co-doping with Si is investigated in a samples series with variable TMGa and Cp$_2$Fe. EXAFS spectra as a function of both TMGa and Cp$_2$Fe are reported in Fig.~\ref{fig:gafensi-chi} with their relative FTs in Fig.~\ref{fig:gafensi-r}. By considering the same part of the spectra highlighted in the previous two series, it is visible in this case that no differences emerge between spectra \cite{Note:distortion} and all present a typical Fe$_{\mathrm{Ga}}$ signal. The Fe full inclusion in substitutional sites is confirmed by the quantitative analysis reported in Table~\ref{tab:dists}C.

\subsection{Fe charge state}

 In order to get information on the charge state of Fe in the studied samples, the near-edge region of the absorption spectra (XANES) is analyzed. The amplitude and position of the peaks due to the partially forbidden $1s \to 3d$ transitions appearing in the pre-edge region of the absorption coefficient were widely investigated in literature as a function of the local symmetry (tetrahedral or octahedral) and valence state (Fe$^{3+}$ or Fe$^{2+}$) in Fe compounds.\cite{Galoisy:2001_CG,Westre:1997_JACS} The general finding was that tetrahedrally coordinated compounds exhibit a single pre-edge peak with an amplitude above 10 \% of the total edge jump and a position changing with the valence state from $\approx 7112$~eV in the case of Fe$^{2+}$ to $\approx 7114$~eV in the case of Fe$^{3+}$.\cite{Galoisy:2001_CG}

\begin{figure*}
\includegraphics[width=0.49\textwidth]{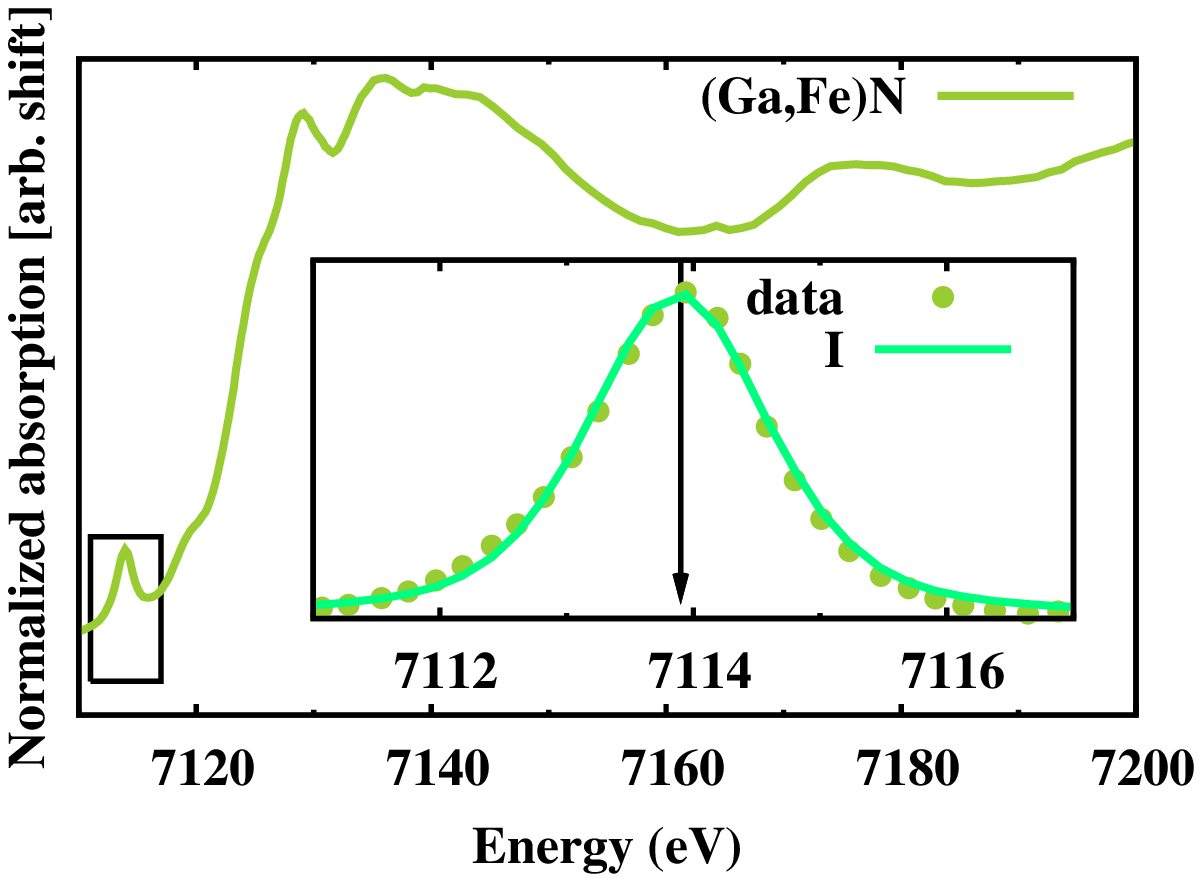} \includegraphics[width=0.49\textwidth]{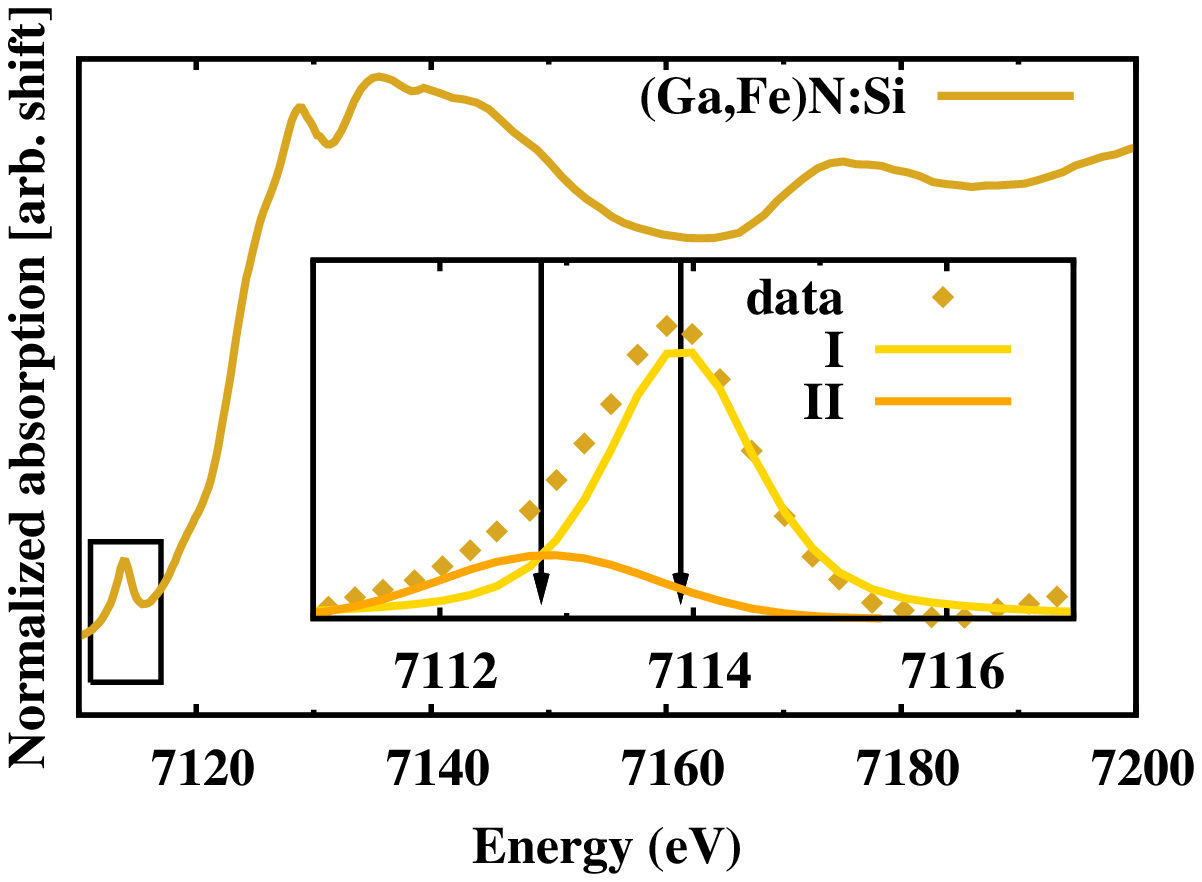}
 \caption{(Color online) Normalized XANES spectra for (Ga,Fe)N [left] and (Ga,Fe)N:Si [right] with the pre-peak reported in the inset after baseline subtraction and the fitted functions with centroids indicated by arrows.}\label{fig:xanes}
\end{figure*}

 Here we present a systematic extension of a previous investigation.\cite{Bonanni:2008_PRL} Quantitative results are reported in Table~\ref{tab:xanes} and two representative fits are plotted in Fig.~\ref{fig:xanes}; undoped samples exhibit only a single peak of average amplitude $13 \pm 1$ \% and positon $7113.9 \pm 0.1$~eV whereas the doped samples present the same peak at $7113.9 \pm 0.1$~eV with a slightly lower average amplitude of $10 \pm 1$ \% and a further peak at $7112.8 \pm 0.1$~eV of amplitude $3 \pm 1$ \%. These data are explained by the presence of Fe$^{3+}$ ions in the first series and a co-existence of Fe$^{3+}$ and Fe$^{2+}$ ions in the second case. The appearance of double peaks in the pre-edge region in case of co-existence of chemical species at different valence states was already pointed out in literature \cite{Giuli:2002_GCA,Westre:1997_JACS} and in our case evidences the partial reduction of the metal ions (Fe$^{3+}$ $\to$ Fe$^{2+}$) upon Si addition.

\begin{table}
 \caption{Quantitative results of the XANES analysis.}\label{tab:xanes}
\begin{tabular}{|cc|cc|cc|}
\hline
\hline
Cp$_2$Fe & TMGa   &  \multicolumn{2}{c|}{ peak I}      & \multicolumn{2}{c|}{ peak II} \\
         &        & center  &  height &  center &  height \\
(sccm)   & (sccm) & (eV)    &  (\%)   &  (eV)   &  (\%)   \\
\hline
\multicolumn{6}{|l|}{(Ga,Fe)N}                                                 \\
\hline
300      & 5      & 7113.8(1)    &  8(1)         &  ---         &  ---         \\
300      & 8      & 7113.9(1)    &  12(1)        &  ---         &  ---         \\
300      & 10     & 7113.9(1)    &  14(1)        &  ---         &  ---         \\
300      & 12     & 7113.9(1)    &  17(1)        &  ---         &  ---         \\
\hline
\multicolumn{6}{|l|}{(Ga,Fe)N:Si}                                              \\
\hline
300      & 5      & 7113.9(1)    &  10(1)        &  7112.9(1)   &  2(1)        \\
300      & 10     & 7113.9(1)    &  11(1)        &  7112.9(1)   &  3(1)        \\
250      & 10     & 7113.9(1)    &  11(1)        &  7112.7(1)   &  3(1)        \\
100      & 10     & 7113.8(1)    &  9(1)         &  7112.5(1)   &  4(1)        \\
\hline
\hline
\end{tabular}
\end{table}

\subsection{Theoretical results}

\subsubsection{Substitutional Fe in GaN}

 The local structure of a Fe$_{\mathrm{Ga}}$ in GaN is investigated through geometry optimization procedures both in the LSD and LSD+U frameworks. The estimated lattice constants of the 72-atoms supercell are $a$ = 9.66~\AA, $c$ = 10.49~\AA. Details of the atomic geometries produced by the (more reliable) LSD+U calculations are given in Table~\ref{tab:dists}E for the charge states 0, +1 and -1 of Fe$_{\mathrm{Ga}}$, which correspond  to the Fe$_{Ga}^{3+}$, Fe$_{Ga}^{4+}$ and Fe$_{Ga}^{2+}$ forms of the impurity, respectively. Present results show that the Fe$_{\mathrm{Ga}}$-N bond distances for the -1 (+1) state are slightly longer (almost the same) with respect to those of the neutral state. Moreover, no significant differences are found between the Fe$_{\mathrm{Ga}}$-Ga distances estimated for the above three charge states of Fe$_{\mathrm{Ga}}$. A quite similar picture is provided by the LSD calculations.

\begin{figure}
\includegraphics[width=.45\textwidth]{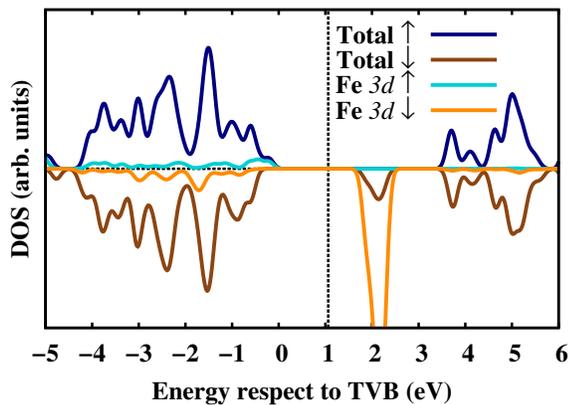}
  \caption{(Color online) LSD+U total spin density of states (DOS) and DOS projected on the Fe atomic orbitals for a substitutional Fe in the neutral charge state. The DOS for Fe are magnified ten times for sake of clarity. The calculated Fermi energy (E$_{\mathrm{F}}$) is 10.52~eV and is indicated by a vertical dashed line to respect to the top of the valence band (TVB).}\label{fig:dos-sfe}
\end{figure}

 Regarding the electronic properties of Fe$_{\mathrm{Ga}}$, total spin DOS and spin DOS projected on the Fe atomic orbitals as given by the LSD+U calculations for the neutral state of the impurity are reported in Fig.~\ref{fig:dos-sfe} together with the calculated Fermi energy. These DOS have to be considered together with an estimate of 1.7~eV given by the Kohn-Sham electronic eigenvalues for the GaN energy gap ($E_{\mathrm{g}}$). Then, Fig.~\ref{fig:dos-sfe} shows that occupied electronic states induced by the substitutional Fe are mixed with GaN states at the top of the valence band (TVB), whereas unoccupied Fe states are resonant with the conduction band. A more reliable estimate of the location of the levels induced by a Fe$_{\mathrm{Ga}}$ in the GaN energy gap is achieved by calculating the corresponding $\epsilon^{0/-1}$ transition level where the Fe impurity changes its charge state from 0 to -1. The present evaluation of the $\epsilon^{0/-1}$ level is 2.25~eV to be compared with a calculated $E_{\mathrm{g}}$ of 2.93~eV. These results favorably compare with experimental results locating the Fe acceptor level at 2.86~eV,\cite{Malguth:2006_PRB} on a band gap of 3.50~eV for wz-GaN at 0~K.\cite{Monemar:1974_PRB} It can be noted that an  even better agreement with the experiment could be reached by assuming  a linear scaling of $\epsilon^{0/-1}$ with the energy gap. In this case, indeed, a value of 2.70~eV for the Fe level would correspond to the experimental energy gap.

 The present results agree with a deep acceptor character of Fe$_{\mathrm{Ga}}$. They seem also to rule out a possible donor behavior of Fe$_{\mathrm{Ga}}$. In Fig.~\ref{fig:dos-sfe} it is shown indeed that there are no occupied Fe states in the GaN energy gap. Moreover, a value of $-0.03$~eV has been estimated for the $\epsilon^{0/+1}$ transition level indicating that a Fe$^{3+/4+}$ donor level of Fe$_{\mathrm{Ga}}$ would be resonant with the valence band. In agreement with this conclusion no such state has been found experimentally in the band gap of GaN:Fe.\cite{Malguth:2008_PSSB} However, the present {\em ab initio} findings do not rule out a possibility that the Fe impurity in GaN gives rise to a charge transfer state, Fe$^{3+} + h$, acting as a hole trap, as postulated by some of us.\cite{Pacuski:2008_PRL,Dietl:2008_PRB}

 The results of the present calculations permit also to asses estimate a change in the formation energies of Fe$_{\mathrm{Ga}}$ in different charge states ($\Omega_{\rm F}^{q}$) as a function of the Fermi energy E$_{\mathrm{F}}$. For instance, for the magnitude of $\Omega_{\rm F}^{-1}$-$\Omega_{\rm F}^{0}$, we estimate  +0.40~eV and -1.35~eV for E$_{\mathrm{F}}$ equal to $E_{\mathrm{g}}/2$ and $E_{\mathrm{g}}$, respectively (in this case we use the experimental value of the energy gap $E_{\mathrm{g}}$). This indicates that the formation of Fe$_{Ga}^{3+}$ is promoted in intrinsic GaN, while Fe$_{Ga}^{2+}$ is favored in the $n$-doped material. On the contrary, the value of $\Omega_{\rm F}^{+1}$-$\Omega_{\rm F}^{0}$ is evaluated as +0.10~eV and +1.85~eV for the Fermi level at the top of the valence band and in the middle of the energy gap, respectively, that is, a donor behavior of Fe (corresponding to the formation of Fe$_{Ga}^{4+}$) is never favored.

\subsubsection{$\epsilon$-Fe$_3$N}

\begin{figure}
\includegraphics[width=.45\textwidth]{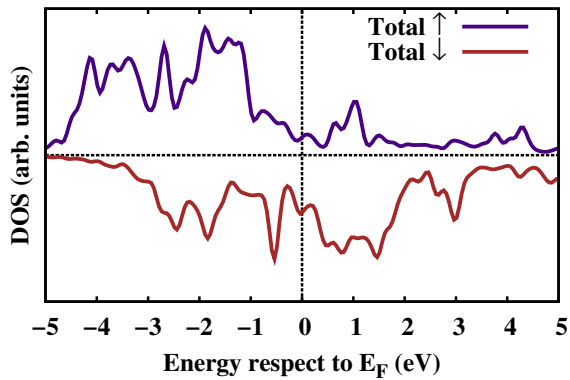}
  \caption{(Color online) LSD total spin density of states (DOS) for $\epsilon$-Fe$_3$N. The calculated Fermi energy (E$_F$) is 17.64~eV.}\label{fig:dos-fen}
\end{figure}

 The cell parameters have been calculated by minimizing the total energy of the system. Optimal values have been found at the zero value of linear fits of the total stress, corresponding to the minimum of the total energy. Such a procedure gives $a$ = 4.64~\AA\ and $c$ = 4.34~\AA, that is $\approx$~2\% contraction of the experimental values $a_{exp}$ = 4.7209(6)~\AA\ and $c_{exp}$ = 4.4188(9)~\AA; this being a typical effect of the LSD approximation. The relaxation properly reproduces the experimental value of $x$ that is a 0.009 contraction from the 1/3 ideal value of $x$ for Fe in the hcp structure. The resulting atomic distances reported in Table~\ref{tab:dists}E are slightly different from the crystallographic ones\cite{Jacobs:1995_JAC} (Table~\ref{tab:dists}D). The above structural minimization corresponds to a magnetization of 2.2 $\mu_B$/atom, in agreement with the experimental value found from neutron diffraction at 4.2~K.\cite{Leineweber:1999_JAC} Finally, the total DOS shown in Fig.~\ref{fig:dos-fen} gives a clear evidence of the metallic behavior of this system. These results are achieved by LSD calculations; the LSD+U scheme is also tested with different U values in the range 2--5~eV and no appreciable difference are found between the LSD and LSD+U results.

\section{Discussion}\label{sec:discussion}

\subsection{EXAFS}

 In the first series presented in this study the sample with a lower Fe content exhibits two coordination shells around Fe. These shells are identified as Fe--N end Fe--Ga contributions and the observed bond length values are in agreement with those calculated via DFT for a Fe ion in a Ga-substitutional site. Thus, we conclude that Fe substitutes Ga, as previously suggested by a qualitative analysis of EXAFS data.\cite{Kumagai:2006_JCG} The closeness of the theoretical estimates of the Fe--N distances reported in Table~\ref{tab:dists}E makes it difficult to exploit the Fe$_{\mathrm{Ga}}$ local geometry to distinguish between its possible, different charge states. However, the theoretical results support a deep acceptor character of Fe$_{\mathrm{Ga}}$ and locate the corresponding electronic level in the upper half of the GaN energy gap. Accordingly, the $\Omega_{\rm F}^{0}$ formation energy is 0.40~eV lower in energy than $\Omega_{\rm F}^{-1}$ in intrinsic GaN, thus suggesting the presence of Fe$_{\mathrm{Ga}}$  in the neutral state.

 For samples grown with higher Fe content an additional shell appears, matching a coordination at $2.70-2.75$~\AA. This value coincides with the experimental and calculated Fe--Fe second shell distance in $\epsilon$-Fe$_3$N. This suggests that Fe participates in the form of $\epsilon$-Fe$_3$N. It is worth noting that in this compound the first shell corresponds to the Fe--N bonds exhibiting the same structural parameters (length and number of neighbors) as those specific to Ga-substitutional Fe in GaN. These results provide an independent confirmation of our previous studies on (Ga,Fe)N,\cite{Bonanni:2007_PRB,Li:2008_JCG} according to which Fe-rich nanocrystals were evidenced by neither HRTEM nor HRXRD and a paramagnetic response was observed by SQUID in samples with a low-Fe content, whereas samples with a high Fe content showed the presence of Fe-rich nanocrystals and a ferromagnetic behavior.

 From a reduction in the amplitude of the dominating Fe--Ga signal relative to the pure substitutional specimen it is possible to estimate the relative content of Fe in the two phases (GaN and $\epsilon$-Fe$_3$N), as shown in Table~\ref{tab:dists}. From these data we infer that the flow-rate of 200 sccm Cp$_2$Fe corresponds to the onset of $\epsilon$-Fe$_3$N precipitation at our growth conditions, though the spinodal decomposition, that is the aggregation of Fe cations without crystallographic phase separation may begin at lower Fe contents. Significantly, no presence of $\epsilon$-Fe$_3$N is detected in samples co-doped with Si. 

 It is worth to underline that no evidence is found for the $\alpha$-Fe phase that should be witnessed by a double Fe--Fe shell at 2.499 and 2.886~\AA.\cite{Swanson:1955_NBoS} Actually, this phase was shown to be present only at the surface of samples grown under particular conditions\cite{Li:2008_JCG} and the overall Fe fraction in this phase presumably lies below the detection limit of EXAFS (about 15\% in the present case). 

 We also observe that our data do not point to the presence of Fe in interstitial sites, at least in samples containing no secondary phases, for which the EXAFS signal is completely reproduced by a simple substitutional model. In the case of samples with precipitates, their contribution could {\em a priori} mask a signal coming from interstitials. This is in contrast to the case of (Ga,Mn)As where evidence for Mn interstitials has been found in EXAFS studies.\cite{Bacewicz:2005_JPCS,D'Acapito:2006_PRB}

 Turning to samples deposited at different growth rates we note that if the Fe content exceeds the previously determined solubility limit, we find again the presence of $\epsilon$-Fe$_3$N nanocrystals in the film grown at the lowest rate (TMGa = 5~sccm). For faster growth rates, however, the EXAFS spectra correspond exclusively to substitutional Fe$_{\mathrm{Ga}}$ in GaN, emphasizing the effectiveness of the fast growth in suppressing the formation of segregated phases. This observation is in qualitative agreement with the previous study,\cite{Navarro-Quezada:2008_JCG} where HRTEM carried out for samples grown at low and high rates (although at a noticeably higher Fe content) revealed the formation of $\epsilon$-Fe$_3$N only for slowly grown specimens.

\subsection{XANES}

 The XANES spectral region is characterized by the presence of a single pre-edge peak at about 7114~eV in all samples grown without Si. In agreement with the literature data\cite{Galoisy:2001_CG,Westre:1997_JACS,Giuli:2002_GCA} this reveals that Fe assumes the Fe$^{3+}$ charge state in a tetrahedral environment, and this result corroborates the conclusions derived from previous electron paramagnetic resonance studies\cite{Bonanni:2007_PRB,Malguth:2008_PSSB} and from the present DFT calculation for (Ga,Fe)N. Co-doping with Si results in the appearance of an additional peak in the pre-edge region at about 7112.8~eV, which is attributed to Fe$^{2+}$ ions in a tetrahedral environment, as already pointed out in our prior work.\cite{Bonanni:2008_PRL} The fraction of the total Fe content in this particular valence state is below 20 \%, the majority of Fe impurities still remaining in the Fe$^{3+}$ configuration. This explains why we do not observe in the EXAFS data (and, in particular, in the values of the Fe--N bond length) any significant deviation when comparing the samples containing no Si impurities with the Si-doped ones. However, the co-deposition of Si hampers the aggregation of Fe and, thus, shifts the solubility limit to higher Fe concentrations.

 The above interpretation of the XANES findings is consistent with the {\em ab initio} results, implying that the Fe$^{2+}$/Fe$^{3+}$ state, i.e. the  $\epsilon^{0/-1}$ level, resides in the GaN gap 2.25~eV above the top of the valence band. Thus, this state should be occupied by electrons provided by shallow donors, such as Si. This conclusion is further supported by the computed formation energy that results to be for the Fe$^{3+}$ configuration by 1.35~eV higher than for the Fe$^{2+}$ case.

\section{Conclusions and outlook}\label{sec:conclusions}

 The present EXAFS results, together with previous HRTEM, electron dispersive spectroscopy, and synchrotron XRD,\cite{Bonanni:2008_PRL} show how the Fe incorporation can be efficiently controlled by Fe flow, growth rate, and co-doping with Si. In particular, the aggregation of Fe cations can be minimized by increasing the growth rate and by co-doping with Si, shifting the solubility limit towards higher Fe content at given growth conditions. While there is agreement between the present DFT computations and both the previously \cite{Bonanni:2007_PRB,Malguth:2008_PSSB} and here observed Fe charge state, including its evolution with the Si concentration, a detailed understanding of how co-doping and other growth parameters affect the aggregation of Fe ions awaits for a comprehensive theoretical treatment.

 A strict correlation between magnetic properties and the presence of Fe-rich nanocrystals \cite{Bonanni:2007_PRB,Bonanni:2008_PRL} strongly suggests that the surprisingly robust ferromagnetism of (Ga,Fe)N as well as of other diluted magnetic semiconductors and oxides deposited under specific growth conditions results from the self-organized assembly of magnetic nanocrystals, whose high blocking temperatures account for the survival of ferromagnetic features at high temperatures. This emerging insight provides a promising starting point for exploiting a number of expected functionalities of these nanocomposite semiconductor/ferromagnetic metal systems.\cite{Katayama-Yoshida:2007_PSSA,Dietl:2008_JAP}.

\section{Acknowledgements}

 We acknowledge the European Synchrotron Radiation Facility and the Italian Collaborating Research Group for the provision of the synchrotron radiation facilities. GILDA is a project jointly financed by CNR and INFN. M.~R. is indebted to Piotr Bogus{\l}awski and Pawe{\l} Jakubas for valuable discussions during his visit in Warsaw and to CNR for financial support by the Short Term Mobility 2008 program (N.0002372). We also acknowledge Federico Boscherini for fruitful discussions. This work has been partly supported by the Austrian Fonds zur {F\"{o}rderung} der wissenschaftlichen Forschung (P18942, P20065 and N107-NAN), by the ESF FoNe project SPINTRA (ERAS-CT-2003-980409), and by the FunDMS Advanced Grant within the European Research Council ``Ideas'' Programme of EC 7FP.



\begin{thebibliography}{56}
\expandafter\ifx\csname natexlab\endcsname\relax\def\natexlab#1{#1}\fi
\expandafter\ifx\csname bibnamefont\endcsname\relax
  \def\bibnamefont#1{#1}\fi
\expandafter\ifx\csname bibfnamefont\endcsname\relax
  \def\bibfnamefont#1{#1}\fi
\expandafter\ifx\csname citenamefont\endcsname\relax
  \def\citenamefont#1{#1}\fi
\expandafter\ifx\csname url\endcsname\relax
  \def\url#1{\texttt{#1}}\fi
\expandafter\ifx\csname urlprefix\endcsname\relax\def\urlprefix{URL }\fi
\providecommand{\bibinfo}[2]{#2}
\providecommand{\eprint}[2][]{\url{#2}}

\bibitem[{\citenamefont{Kuroda et~al.}(2007)\citenamefont{Kuroda, Nishizawa,
  Takita, Mitome, Bando, Osuch, and Dietl}}]{Kuroda:2007_NM}
\bibinfo{author}{\bibfnamefont{S.}~\bibnamefont{Kuroda}},
  \bibinfo{author}{\bibfnamefont{N.}~\bibnamefont{Nishizawa}},
  \bibinfo{author}{\bibfnamefont{K.}~\bibnamefont{Takita}},
  \bibinfo{author}{\bibfnamefont{M.}~\bibnamefont{Mitome}},
  \bibinfo{author}{\bibfnamefont{Y.}~\bibnamefont{Bando}},
  \bibinfo{author}{\bibfnamefont{K.}~\bibnamefont{Osuch}}, \bibnamefont{and}
  \bibinfo{author}{\bibfnamefont{T.}~\bibnamefont{Dietl}},
  \bibinfo{journal}{Nat. Mater.} \textbf{\bibinfo{volume}{6}},
  \bibinfo{pages}{440} (\bibinfo{year}{2007}).

\bibitem[{\citenamefont{Chambers et~al.}(2006)\citenamefont{Chambers, Droubay,
  Wang, Rosso, Heald, Schwartz, Kittilstved, and Gamelin}}]{Chambers:2006_MT}
\bibinfo{author}{\bibfnamefont{S.}~\bibnamefont{Chambers}},
  \bibinfo{author}{\bibfnamefont{T.}~\bibnamefont{Droubay}},
  \bibinfo{author}{\bibfnamefont{C.}~\bibnamefont{Wang}},
  \bibinfo{author}{\bibfnamefont{K.}~\bibnamefont{Rosso}},
  \bibinfo{author}{\bibfnamefont{S.}~\bibnamefont{Heald}},
  \bibinfo{author}{\bibfnamefont{D.}~\bibnamefont{Schwartz}},
  \bibinfo{author}{\bibfnamefont{K.}~\bibnamefont{Kittilstved}},
  \bibnamefont{and} \bibinfo{author}{\bibfnamefont{D.}~\bibnamefont{Gamelin}},
  \bibinfo{journal}{Materials Today} \textbf{\bibinfo{volume}{9}},
  \bibinfo{pages}{28} (\bibinfo{year}{2006}).

\bibitem[{\citenamefont{Kobayashi et~al.}(2008)\citenamefont{Kobayashi, Ishida,
  Hwang, Song, Fujimori, Yang, Lee, Lin, Huang, Chen
  et~al.}}]{Kobayashi:2008_NJP}
\bibinfo{author}{\bibfnamefont{M.}~\bibnamefont{Kobayashi}},
  \bibinfo{author}{\bibfnamefont{Y.}~\bibnamefont{Ishida}},
  \bibinfo{author}{\bibfnamefont{J.~I.} \bibnamefont{Hwang}},
  \bibinfo{author}{\bibfnamefont{G.~S.} \bibnamefont{Song}},
  \bibinfo{author}{\bibfnamefont{A.}~\bibnamefont{Fujimori}},
  \bibinfo{author}{\bibfnamefont{C.~S.} \bibnamefont{Yang}},
  \bibinfo{author}{\bibfnamefont{L.}~\bibnamefont{Lee}},
  \bibinfo{author}{\bibfnamefont{H.-J.} \bibnamefont{Lin}},
  \bibinfo{author}{\bibfnamefont{D.~J.} \bibnamefont{Huang}},
  \bibinfo{author}{\bibfnamefont{C.~T.} \bibnamefont{Chen}},
  \bibnamefont{et~al.}, \bibinfo{journal}{New J. Phys.}
  \textbf{\bibinfo{volume}{10}}, \bibinfo{pages}{055011}
  (\bibinfo{year}{2008}).

\bibitem[{\citenamefont{Bonanni}(2007)}]{Bonanni:2007_SST}
\bibinfo{author}{\bibfnamefont{A.}~\bibnamefont{Bonanni}},
  \bibinfo{journal}{Semicond. Sci. Technol.} \textbf{\bibinfo{volume}{22}},
  \bibinfo{pages}{R41} (\bibinfo{year}{2007}), \bibinfo{note}{and references
  therein}.

\bibitem[{\citenamefont{Dietl}(2007)}]{Dietl:2007_JPCM}
\bibinfo{author}{\bibfnamefont{T.}~\bibnamefont{Dietl}}, \bibinfo{journal}{J.
  Phys.: Condens. Matter} \textbf{\bibinfo{volume}{19}},
  \bibinfo{pages}{165204} (\bibinfo{year}{2007}).

\bibitem[{\citenamefont{Katayama-Yoshida
  et~al.}(2007)\citenamefont{Katayama-Yoshida, Sato, Fukushima, Toyoda, Kizaki,
  Dinh, and Dederichs}}]{Katayama-Yoshida:2007_PSSA}
\bibinfo{author}{\bibfnamefont{H.}~\bibnamefont{Katayama-Yoshida}},
  \bibinfo{author}{\bibfnamefont{K.}~\bibnamefont{Sato}},
  \bibinfo{author}{\bibfnamefont{T.}~\bibnamefont{Fukushima}},
  \bibinfo{author}{\bibfnamefont{M.}~\bibnamefont{Toyoda}},
  \bibinfo{author}{\bibfnamefont{H.}~\bibnamefont{Kizaki}},
  \bibinfo{author}{\bibfnamefont{V.~A.} \bibnamefont{Dinh}}, \bibnamefont{and}
  \bibinfo{author}{\bibfnamefont{P.~H.} \bibnamefont{Dederichs}},
  \bibinfo{journal}{Phys. Status Solidi A} \textbf{\bibinfo{volume}{204}},
  \bibinfo{pages}{15} (\bibinfo{year}{2007}).

\bibitem[{\citenamefont{Bonanni et~al.}(2007)\citenamefont{Bonanni, Kiecana,
  Simbrunner, Li, Sawicki, Wegscheider, Quast, Przybylinska, Navarro-Quezada,
  Jakiela et~al.}}]{Bonanni:2007_PRB}
\bibinfo{author}{\bibfnamefont{A.}~\bibnamefont{Bonanni}},
  \bibinfo{author}{\bibfnamefont{M.}~\bibnamefont{Kiecana}},
  \bibinfo{author}{\bibfnamefont{C.}~\bibnamefont{Simbrunner}},
  \bibinfo{author}{\bibfnamefont{T.}~\bibnamefont{Li}},
  \bibinfo{author}{\bibfnamefont{M.}~\bibnamefont{Sawicki}},
  \bibinfo{author}{\bibfnamefont{M.}~\bibnamefont{Wegscheider}},
  \bibinfo{author}{\bibfnamefont{M.}~\bibnamefont{Quast}},
  \bibinfo{author}{\bibfnamefont{H.}~\bibnamefont{Przybylinska}},
  \bibinfo{author}{\bibfnamefont{A.}~\bibnamefont{Navarro-Quezada}},
  \bibinfo{author}{\bibfnamefont{R.}~\bibnamefont{Jakiela}},
  \bibnamefont{et~al.}, \bibinfo{journal}{Phys. Rev. B}
  \textbf{\bibinfo{volume}{75}}, \bibinfo{pages}{125210}
  (\bibinfo{year}{2007}).

\bibitem[{\citenamefont{Pacuski et~al.}(2008)\citenamefont{Pacuski, Kossacki,
  Ferrand, Golnik, Cibert, Wegscheider, Navarro-Quezada, Bonanni, Kiecana,
  Sawicki et~al.}}]{Pacuski:2008_PRL}
\bibinfo{author}{\bibfnamefont{W.}~\bibnamefont{Pacuski}},
  \bibinfo{author}{\bibfnamefont{P.}~\bibnamefont{Kossacki}},
  \bibinfo{author}{\bibfnamefont{D.}~\bibnamefont{Ferrand}},
  \bibinfo{author}{\bibfnamefont{A.}~\bibnamefont{Golnik}},
  \bibinfo{author}{\bibfnamefont{J.}~\bibnamefont{Cibert}},
  \bibinfo{author}{\bibfnamefont{M.}~\bibnamefont{Wegscheider}},
  \bibinfo{author}{\bibfnamefont{A.}~\bibnamefont{Navarro-Quezada}},
  \bibinfo{author}{\bibfnamefont{A.}~\bibnamefont{Bonanni}},
  \bibinfo{author}{\bibfnamefont{M.}~\bibnamefont{Kiecana}},
  \bibinfo{author}{\bibfnamefont{M.}~\bibnamefont{Sawicki}},
  \bibnamefont{et~al.}, \bibinfo{journal}{Phys. Rev. Lett.}
  \textbf{\bibinfo{volume}{100}}, \bibinfo{pages}{037204}
  (\bibinfo{year}{2008}).

\bibitem[{\citenamefont{Bonanni et~al.}(2008)\citenamefont{Bonanni,
  Navarro-Quezada, Li, Wegscheider, Matej, Holy, Lechner, Bauer, Rovezzi,
  D'Acapito et~al.}}]{Bonanni:2008_PRL}
\bibinfo{author}{\bibfnamefont{A.}~\bibnamefont{Bonanni}},
  \bibinfo{author}{\bibfnamefont{A.}~\bibnamefont{Navarro-Quezada}},
  \bibinfo{author}{\bibfnamefont{T.}~\bibnamefont{Li}},
  \bibinfo{author}{\bibfnamefont{M.}~\bibnamefont{Wegscheider}},
  \bibinfo{author}{\bibfnamefont{Z.}~\bibnamefont{Matej}},
  \bibinfo{author}{\bibfnamefont{V.}~\bibnamefont{Holy}},
  \bibinfo{author}{\bibfnamefont{R.~T.} \bibnamefont{Lechner}},
  \bibinfo{author}{\bibfnamefont{G.}~\bibnamefont{Bauer}},
  \bibinfo{author}{\bibfnamefont{M.}~\bibnamefont{Rovezzi}},
  \bibinfo{author}{\bibfnamefont{F.}~\bibnamefont{D'Acapito}},
  \bibnamefont{et~al.}, \bibinfo{journal}{Phys. Rev. Lett.}
  \textbf{\bibinfo{volume}{101}}, \bibinfo{pages}{135502}
  (\bibinfo{year}{2008}).

\bibitem[{\citenamefont{Lee et~al.}(1981)\citenamefont{Lee, Citrin,
  Eisenberger, and Kincaid}}]{Lee:1981_RMP}
\bibinfo{author}{\bibfnamefont{P.~A.} \bibnamefont{Lee}},
  \bibinfo{author}{\bibfnamefont{P.~H.} \bibnamefont{Citrin}},
  \bibinfo{author}{\bibfnamefont{P.}~\bibnamefont{Eisenberger}},
  \bibnamefont{and} \bibinfo{author}{\bibfnamefont{B.~M.}
  \bibnamefont{Kincaid}}, \bibinfo{journal}{Rev. Mod. Phys.}
  \textbf{\bibinfo{volume}{53}}, \bibinfo{pages}{769} (\bibinfo{year}{1981}).

\bibitem[{\citenamefont{Boscherini}(2008)}]{Boscherini:2008_book}
\bibinfo{author}{\bibfnamefont{F.}~\bibnamefont{Boscherini}},
  \emph{\bibinfo{title}{X-ray absorption fine structure in the study of
  semiconductor heterostructures and nanostructures}}
  (\bibinfo{publisher}{Elsevier}, \bibinfo{year}{2008}),
  chap.~\bibinfo{chapter}{9}, p. \bibinfo{pages}{289}.

\bibitem[{\citenamefont{Soo et~al.}(2001{\natexlab{a}})\citenamefont{Soo,
  Kioseoglou, Huang, Kim, Kao, Takatani, Haneda, and Munekata}}]{Soo:2001_JSR}
\bibinfo{author}{\bibfnamefont{Y.~L.} \bibnamefont{Soo}},
  \bibinfo{author}{\bibfnamefont{G.}~\bibnamefont{Kioseoglou}},
  \bibinfo{author}{\bibfnamefont{S.}~\bibnamefont{Huang}},
  \bibinfo{author}{\bibfnamefont{S.}~\bibnamefont{Kim}},
  \bibinfo{author}{\bibfnamefont{Y.~H.} \bibnamefont{Kao}},
  \bibinfo{author}{\bibfnamefont{Y.}~\bibnamefont{Takatani}},
  \bibinfo{author}{\bibfnamefont{S.}~\bibnamefont{Haneda}}, \bibnamefont{and}
  \bibinfo{author}{\bibfnamefont{H.}~\bibnamefont{Munekata}},
  \bibinfo{journal}{J. Synchrotron Rad.} \textbf{\bibinfo{volume}{8}},
  \bibinfo{pages}{874} (\bibinfo{year}{2001}{\natexlab{a}}).

\bibitem[{\citenamefont{Soo et~al.}(2001{\natexlab{b}})\citenamefont{Soo,
  Kioseoglou, Kim, Huang, Kao, Kuwabara, Owa, Kondo, and
  Munekata}}]{Soo:2001_APL}
\bibinfo{author}{\bibfnamefont{Y.~L.} \bibnamefont{Soo}},
  \bibinfo{author}{\bibfnamefont{G.}~\bibnamefont{Kioseoglou}},
  \bibinfo{author}{\bibfnamefont{S.}~\bibnamefont{Kim}},
  \bibinfo{author}{\bibfnamefont{S.}~\bibnamefont{Huang}},
  \bibinfo{author}{\bibfnamefont{Y.~H.} \bibnamefont{Kao}},
  \bibinfo{author}{\bibfnamefont{S.}~\bibnamefont{Kuwabara}},
  \bibinfo{author}{\bibfnamefont{S.}~\bibnamefont{Owa}},
  \bibinfo{author}{\bibfnamefont{T.}~\bibnamefont{Kondo}}, \bibnamefont{and}
  \bibinfo{author}{\bibfnamefont{H.}~\bibnamefont{Munekata}},
  \bibinfo{journal}{Appl. Phys. Lett.} \textbf{\bibinfo{volume}{79}},
  \bibinfo{pages}{3926} (\bibinfo{year}{2001}{\natexlab{b}}).

\bibitem[{\citenamefont{Soo et~al.}(2004)\citenamefont{Soo, Kim, Kao, Blattner,
  Wessels, Khalid, Hanke, and Kao}}]{Soo:2004_APL}
\bibinfo{author}{\bibfnamefont{Y.~L.} \bibnamefont{Soo}},
  \bibinfo{author}{\bibfnamefont{S.}~\bibnamefont{Kim}},
  \bibinfo{author}{\bibfnamefont{Y.~H.} \bibnamefont{Kao}},
  \bibinfo{author}{\bibfnamefont{A.~J.} \bibnamefont{Blattner}},
  \bibinfo{author}{\bibfnamefont{B.~W.} \bibnamefont{Wessels}},
  \bibinfo{author}{\bibfnamefont{S.}~\bibnamefont{Khalid}},
  \bibinfo{author}{\bibfnamefont{C.~S.} \bibnamefont{Hanke}}, \bibnamefont{and}
  \bibinfo{author}{\bibfnamefont{C.-C.} \bibnamefont{Kao}},
  \bibinfo{journal}{Appl. Phys. Lett.} \textbf{\bibinfo{volume}{84}},
  \bibinfo{pages}{481} (\bibinfo{year}{2004}).

\bibitem[{\citenamefont{D'Acapito et~al.}(2006)\citenamefont{D'Acapito,
  Smolentsev, Boscherini, Piccin, Bais, Rubini, Martelli, and
  Franciosi}}]{D'Acapito:2006_PRB}
\bibinfo{author}{\bibfnamefont{F.}~\bibnamefont{D'Acapito}},
  \bibinfo{author}{\bibfnamefont{G.}~\bibnamefont{Smolentsev}},
  \bibinfo{author}{\bibfnamefont{F.}~\bibnamefont{Boscherini}},
  \bibinfo{author}{\bibfnamefont{M.}~\bibnamefont{Piccin}},
  \bibinfo{author}{\bibfnamefont{G.}~\bibnamefont{Bais}},
  \bibinfo{author}{\bibfnamefont{S.}~\bibnamefont{Rubini}},
  \bibinfo{author}{\bibfnamefont{F.}~\bibnamefont{Martelli}}, \bibnamefont{and}
  \bibinfo{author}{\bibfnamefont{A.}~\bibnamefont{Franciosi}},
  \bibinfo{journal}{Phys. Rev. B} \textbf{\bibinfo{volume}{73}},
  \bibinfo{pages}{035314} (\bibinfo{year}{2006}).

\bibitem[{\citenamefont{Bacewicz et~al.}(2005)\citenamefont{Bacewicz, Twar\'og,
  Malinowska, Wojtowicz, Liu, and Furdyna}}]{Bacewicz:2005_JPCS}
\bibinfo{author}{\bibfnamefont{R.}~\bibnamefont{Bacewicz}},
  \bibinfo{author}{\bibfnamefont{A.}~\bibnamefont{Twar\'og}},
  \bibinfo{author}{\bibfnamefont{A.}~\bibnamefont{Malinowska}},
  \bibinfo{author}{\bibfnamefont{T.}~\bibnamefont{Wojtowicz}},
  \bibinfo{author}{\bibfnamefont{X.}~\bibnamefont{Liu}}, \bibnamefont{and}
  \bibinfo{author}{\bibfnamefont{J.}~\bibnamefont{Furdyna}},
  \bibinfo{journal}{J. Phys. Chem. Solids} \textbf{\bibinfo{volume}{66}},
  \bibinfo{pages}{2004} (\bibinfo{year}{2005}).

\bibitem[{\citenamefont{Rovezzi et~al.}(2008)\citenamefont{Rovezzi, Devillers,
  Arras, d'Acapito, Barski, Jamet, and Pochet}}]{Rovezzi:2008_APL}
\bibinfo{author}{\bibfnamefont{M.}~\bibnamefont{Rovezzi}},
  \bibinfo{author}{\bibfnamefont{T.}~\bibnamefont{Devillers}},
  \bibinfo{author}{\bibfnamefont{E.}~\bibnamefont{Arras}},
  \bibinfo{author}{\bibfnamefont{F.}~\bibnamefont{d'Acapito}},
  \bibinfo{author}{\bibfnamefont{A.}~\bibnamefont{Barski}},
  \bibinfo{author}{\bibfnamefont{M.}~\bibnamefont{Jamet}}, \bibnamefont{and}
  \bibinfo{author}{\bibfnamefont{P.}~\bibnamefont{Pochet}},
  \bibinfo{journal}{Appl. Phys. Lett.} \textbf{\bibinfo{volume}{92}},
  \bibinfo{pages}{242510} (\bibinfo{year}{2008}).

\bibitem[{\citenamefont{Dietl}(2006)}]{Dietl:2006_NM}
\bibinfo{author}{\bibfnamefont{T.}~\bibnamefont{Dietl}}, \bibinfo{journal}{Nat.
  Mater.} \textbf{\bibinfo{volume}{5}}, \bibinfo{pages}{673}
  (\bibinfo{year}{2006}).

\bibitem[{\citenamefont{D'Acapito et~al.}(1998)\citenamefont{D'Acapito,
  Colonna, Pascarelli, Antonioli, Balerna, Bazzini, Boscherini, Campolungo,
  Chini, Dalba et~al.}}]{D'Acapito:1998_EN}
\bibinfo{author}{\bibfnamefont{F.}~\bibnamefont{D'Acapito}},
  \bibinfo{author}{\bibfnamefont{S.}~\bibnamefont{Colonna}},
  \bibinfo{author}{\bibfnamefont{S.}~\bibnamefont{Pascarelli}},
  \bibinfo{author}{\bibfnamefont{G.}~\bibnamefont{Antonioli}},
  \bibinfo{author}{\bibfnamefont{A.}~\bibnamefont{Balerna}},
  \bibinfo{author}{\bibfnamefont{A.}~\bibnamefont{Bazzini}},
  \bibinfo{author}{\bibfnamefont{F.}~\bibnamefont{Boscherini}},
  \bibinfo{author}{\bibfnamefont{F.}~\bibnamefont{Campolungo}},
  \bibinfo{author}{\bibfnamefont{G.}~\bibnamefont{Chini}},
  \bibinfo{author}{\bibfnamefont{G.}~\bibnamefont{Dalba}},
  \bibnamefont{et~al.}, \bibinfo{journal}{ESRF Newsletter}
  \textbf{\bibinfo{volume}{30}}, \bibinfo{pages}{42} (\bibinfo{year}{1998}).

\bibitem[{\citenamefont{Pascarelli et~al.}(1996)\citenamefont{Pascarelli,
  Boscherini, D'Acapito, Hrdy, Meneghini, and Mobilio}}]{Pascarelli:1996_JSR}
\bibinfo{author}{\bibfnamefont{S.}~\bibnamefont{Pascarelli}},
  \bibinfo{author}{\bibfnamefont{F.}~\bibnamefont{Boscherini}},
  \bibinfo{author}{\bibfnamefont{F.}~\bibnamefont{D'Acapito}},
  \bibinfo{author}{\bibfnamefont{J.}~\bibnamefont{Hrdy}},
  \bibinfo{author}{\bibfnamefont{C.}~\bibnamefont{Meneghini}},
  \bibnamefont{and} \bibinfo{author}{\bibfnamefont{S.}~\bibnamefont{Mobilio}},
  \bibinfo{journal}{J. Synchrotron Rad.} \textbf{\bibinfo{volume}{3}},
  \bibinfo{pages}{147} (\bibinfo{year}{1996}).

\bibitem[{\citenamefont{Tullio et~al.}(2001)\citenamefont{Tullio, D'Anca,
  Campolungo, D'Acapito, Boscherini, and Mobilio}}]{Tullio:2001_tech}
\bibinfo{author}{\bibfnamefont{V.}~\bibnamefont{Tullio}},
  \bibinfo{author}{\bibfnamefont{F.}~\bibnamefont{D'Anca}},
  \bibinfo{author}{\bibfnamefont{F.}~\bibnamefont{Campolungo}},
  \bibinfo{author}{\bibfnamefont{F.}~\bibnamefont{D'Acapito}},
  \bibinfo{author}{\bibfnamefont{F.}~\bibnamefont{Boscherini}},
  \bibnamefont{and} \bibinfo{author}{\bibfnamefont{S.}~\bibnamefont{Mobilio}},
  \bibinfo{type}{Internal note}, \bibinfo{institution}{LNF-INFN}
  (\bibinfo{year}{2001}).

\bibitem[{\citenamefont{Bearden and Burr}(1967)}]{Bearden:1967_RMP}
\bibinfo{author}{\bibfnamefont{J.~A.} \bibnamefont{Bearden}} \bibnamefont{and}
  \bibinfo{author}{\bibfnamefont{A.~F.} \bibnamefont{Burr}},
  \bibinfo{journal}{Rev. Mod. Phys.} \textbf{\bibinfo{volume}{39}},
  \bibinfo{pages}{125} (\bibinfo{year}{1967}).

\bibitem[{\citenamefont{Klementev}(2001)}]{Klementev:2001_JPDAP}
\bibinfo{author}{\bibfnamefont{K.~V.} \bibnamefont{Klementev}},
  \bibinfo{journal}{J. Phys. D: Appl. Phys.} \textbf{\bibinfo{volume}{34}},
  \bibinfo{pages}{209} (\bibinfo{year}{2001}).

\bibitem[{\citenamefont{Newville}(2001)}]{Newville:2001_JSR}
\bibinfo{author}{\bibfnamefont{M.}~\bibnamefont{Newville}},
  \bibinfo{journal}{J. Synchrotron Rad.} \textbf{\bibinfo{volume}{8}},
  \bibinfo{pages}{322} (\bibinfo{year}{2001}).

\bibitem[{\citenamefont{Ravel and Newville}(2005)}]{Ravel:2005_JSR}
\bibinfo{author}{\bibfnamefont{B.}~\bibnamefont{Ravel}} \bibnamefont{and}
  \bibinfo{author}{\bibfnamefont{M.}~\bibnamefont{Newville}},
  \bibinfo{journal}{J. Synchrotron Rad.} \textbf{\bibinfo{volume}{12}},
  \bibinfo{pages}{537} (\bibinfo{year}{2005}).

\bibitem[{\citenamefont{Ankudinov et~al.}(1998)\citenamefont{Ankudinov, Ravel,
  Rehr, and Conradson}}]{Ankudinov:1998_PRB}
\bibinfo{author}{\bibfnamefont{A.~L.} \bibnamefont{Ankudinov}},
  \bibinfo{author}{\bibfnamefont{B.}~\bibnamefont{Ravel}},
  \bibinfo{author}{\bibfnamefont{J.~J.} \bibnamefont{Rehr}}, \bibnamefont{and}
  \bibinfo{author}{\bibfnamefont{S.~D.} \bibnamefont{Conradson}},
  \bibinfo{journal}{Phys. Rev. B} \textbf{\bibinfo{volume}{58}},
  \bibinfo{pages}{7565} (\bibinfo{year}{1998}).

\bibitem[{\citenamefont{Decremps et~al.}(2003)\citenamefont{Decremps, Datchi,
  Saitta, Polian, Pascarelli, Di~Cicco, Iti\'e, and
  Baudelet}}]{Decremps:2003_PRB}
\bibinfo{author}{\bibfnamefont{F.}~\bibnamefont{Decremps}},
  \bibinfo{author}{\bibfnamefont{F.}~\bibnamefont{Datchi}},
  \bibinfo{author}{\bibfnamefont{A.~M.} \bibnamefont{Saitta}},
  \bibinfo{author}{\bibfnamefont{A.}~\bibnamefont{Polian}},
  \bibinfo{author}{\bibfnamefont{S.}~\bibnamefont{Pascarelli}},
  \bibinfo{author}{\bibfnamefont{A.}~\bibnamefont{Di~Cicco}},
  \bibinfo{author}{\bibfnamefont{J.~P.} \bibnamefont{Iti\'e}},
  \bibnamefont{and} \bibinfo{author}{\bibfnamefont{F.}~\bibnamefont{Baudelet}},
  \bibinfo{journal}{Phys. Rev. B} \textbf{\bibinfo{volume}{68}},
  \bibinfo{pages}{104101} (\bibinfo{year}{2003}).

\bibitem[{\citenamefont{Wojdyr}()}]{Wojdyr:_misc}
\bibinfo{author}{\bibfnamefont{M.}~\bibnamefont{Wojdyr}}, \bibinfo{note}{see
  {\tt http://www.unipress.waw.pl/fityk/}}.

\bibitem[{Not({\natexlab{a}})}]{Note:atan}
\bibinfo{note}{The function has been parametrized as: $f(x) = a_1
  \mathrm{atan}[(x-a_2) a_3] + a_4$, where $a_1$, $a_2$, $a_3$, $a_4$ are
  fitted with a Levenberg-Marquardt algorithm.}

\bibitem[{\citenamefont{Galoisy et~al.}(2001)\citenamefont{Galoisy, Calas, and
  Arrio}}]{Galoisy:2001_CG}
\bibinfo{author}{\bibfnamefont{L.}~\bibnamefont{Galoisy}},
  \bibinfo{author}{\bibfnamefont{G.}~\bibnamefont{Calas}}, \bibnamefont{and}
  \bibinfo{author}{\bibfnamefont{M.~A.} \bibnamefont{Arrio}},
  \bibinfo{journal}{Chem. Geol.} \textbf{\bibinfo{volume}{174}},
  \bibinfo{pages}{307} (\bibinfo{year}{2001}).

\bibitem[{\citenamefont{Westre et~al.}(1997)\citenamefont{Westre, Kennepohl,
  DeWitt, Hedman, Hodgson, and Solomon}}]{Westre:1997_JACS}
\bibinfo{author}{\bibfnamefont{T.~E.} \bibnamefont{Westre}},
  \bibinfo{author}{\bibfnamefont{P.}~\bibnamefont{Kennepohl}},
  \bibinfo{author}{\bibfnamefont{J.~G.} \bibnamefont{DeWitt}},
  \bibinfo{author}{\bibfnamefont{B.}~\bibnamefont{Hedman}},
  \bibinfo{author}{\bibfnamefont{K.~O.} \bibnamefont{Hodgson}},
  \bibnamefont{and} \bibinfo{author}{\bibfnamefont{E.~I.}
  \bibnamefont{Solomon}}, \bibinfo{journal}{J. Am. Chem. Soc.}
  \textbf{\bibinfo{volume}{119}}, \bibinfo{pages}{6297} (\bibinfo{year}{1997}).

\bibitem[{\citenamefont{Perdew et~al.}(1996)\citenamefont{Perdew, Burke, and
  Ernzerhof}}]{Perdew:1996_PRL}
\bibinfo{author}{\bibfnamefont{J.~P.} \bibnamefont{Perdew}},
  \bibinfo{author}{\bibfnamefont{K.}~\bibnamefont{Burke}}, \bibnamefont{and}
  \bibinfo{author}{\bibfnamefont{M.}~\bibnamefont{Ernzerhof}},
  \bibinfo{journal}{Phys. Rev. Lett.} \textbf{\bibinfo{volume}{77}},
  \bibinfo{pages}{3865} (\bibinfo{year}{1996}).

\bibitem[{\citenamefont{Anisimov et~al.}(1997)\citenamefont{Anisimov,
  Aryasetiawan, and Lichtenstein}}]{Anisimov:1997_JPCM}
\bibinfo{author}{\bibfnamefont{V.~I.} \bibnamefont{Anisimov}},
  \bibinfo{author}{\bibfnamefont{F.}~\bibnamefont{Aryasetiawan}},
  \bibnamefont{and} \bibinfo{author}{\bibfnamefont{A.~I.}
  \bibnamefont{Lichtenstein}}, \bibinfo{journal}{J. Phys.: Condens. Matter}
  \textbf{\bibinfo{volume}{9}}, \bibinfo{pages}{767} (\bibinfo{year}{1997}).

\bibitem[{\citenamefont{Cococcioni and
  de~Gironcoli}(2005)}]{Cococcioni:2005_PRB}
\bibinfo{author}{\bibfnamefont{M.}~\bibnamefont{Cococcioni}} \bibnamefont{and}
  \bibinfo{author}{\bibfnamefont{S.}~\bibnamefont{de~Gironcoli}},
  \bibinfo{journal}{Phys. Rev. B} \textbf{\bibinfo{volume}{71}},
  \bibinfo{pages}{035105} (\bibinfo{year}{2005}).

\bibitem[{\citenamefont{Kulik et~al.}(2006)\citenamefont{Kulik, Cococcioni,
  Scherlis, and Marzari}}]{Kulik:2006_PRL}
\bibinfo{author}{\bibfnamefont{H.~J.} \bibnamefont{Kulik}},
  \bibinfo{author}{\bibfnamefont{M.}~\bibnamefont{Cococcioni}},
  \bibinfo{author}{\bibfnamefont{D.~A.} \bibnamefont{Scherlis}},
  \bibnamefont{and} \bibinfo{author}{\bibfnamefont{N.}~\bibnamefont{Marzari}},
  \bibinfo{journal}{Phys. Rev. Lett.} \textbf{\bibinfo{volume}{97}},
  \bibinfo{pages}{103001} (\bibinfo{year}{2006}).

\bibitem[{\citenamefont{Giannozzi et~al.}()\citenamefont{Giannozzi, Baroni,
  Corso, de~Gironcoli, Cavazzoni, Ballabio, Scandolo, Chiarotti, Focher,
  Pasquarello et~al.}}]{Giannozzi:_misc}
\bibinfo{author}{\bibfnamefont{P.}~\bibnamefont{Giannozzi}},
  \bibinfo{author}{\bibfnamefont{S.}~\bibnamefont{Baroni}},
  \bibinfo{author}{\bibfnamefont{A.~D.} \bibnamefont{Corso}},
  \bibinfo{author}{\bibfnamefont{S.}~\bibnamefont{de~Gironcoli}},
  \bibinfo{author}{\bibfnamefont{C.}~\bibnamefont{Cavazzoni}},
  \bibinfo{author}{\bibfnamefont{G.}~\bibnamefont{Ballabio}},
  \bibinfo{author}{\bibfnamefont{S.}~\bibnamefont{Scandolo}},
  \bibinfo{author}{\bibfnamefont{G.}~\bibnamefont{Chiarotti}},
  \bibinfo{author}{\bibfnamefont{P.}~\bibnamefont{Focher}},
  \bibinfo{author}{\bibfnamefont{A.}~\bibnamefont{Pasquarello}},
  \bibnamefont{et~al.}, \bibinfo{note}{see {\tt
  http://www.quantum-espresso.org}}.

\bibitem[{\citenamefont{Vanderbilt}(1990)}]{Vanderbilt:1990_PRB}
\bibinfo{author}{\bibfnamefont{D.}~\bibnamefont{Vanderbilt}},
  \bibinfo{journal}{Phys. Rev. B} \textbf{\bibinfo{volume}{41}},
  \bibinfo{pages}{7892} (\bibinfo{year}{1990}).

\bibitem[{\citenamefont{Bonapasta et~al.}(2003)\citenamefont{Bonapasta,
  Filippone, and Giannozzi}}]{Bonapasta:2003_PRB}
\bibinfo{author}{\bibfnamefont{A.~A.} \bibnamefont{Bonapasta}},
  \bibinfo{author}{\bibfnamefont{F.}~\bibnamefont{Filippone}},
  \bibnamefont{and}
  \bibinfo{author}{\bibfnamefont{P.}~\bibnamefont{Giannozzi}},
  \bibinfo{journal}{Phys. Rev. B} \textbf{\bibinfo{volume}{68}},
  \bibinfo{pages}{115202} (\bibinfo{year}{2003}).

\bibitem[{\citenamefont{de~Walle and Neugebauer}(2004)}]{Walle:2004_JAP}
\bibinfo{author}{\bibfnamefont{C.~G.~V.} \bibnamefont{de~Walle}}
  \bibnamefont{and}
  \bibinfo{author}{\bibfnamefont{J.}~\bibnamefont{Neugebauer}},
  \bibinfo{journal}{J. Appl. Phys.} \textbf{\bibinfo{volume}{95}},
  \bibinfo{pages}{3851} (\bibinfo{year}{2004}).

\bibitem[{\citenamefont{Amore~Bonapasta
  et~al.}(2004)\citenamefont{Amore~Bonapasta, Filippone, and
  Giannozzi}}]{Amore:2004_PRB}
\bibinfo{author}{\bibfnamefont{A.}~\bibnamefont{Amore~Bonapasta}},
  \bibinfo{author}{\bibfnamefont{F.}~\bibnamefont{Filippone}},
  \bibnamefont{and}
  \bibinfo{author}{\bibfnamefont{P.}~\bibnamefont{Giannozzi}},
  \bibinfo{journal}{Phys. Rev. B} \textbf{\bibinfo{volume}{69}},
  \bibinfo{pages}{115207} (\bibinfo{year}{2004}).

\bibitem[{\citenamefont{Jack}(1952)}]{Jack:1952_AC}
\bibinfo{author}{\bibfnamefont{K.~H.} \bibnamefont{Jack}},
  \bibinfo{journal}{Acta Crystallographica} \textbf{\bibinfo{volume}{5}},
  \bibinfo{pages}{404} (\bibinfo{year}{1952}).

\bibitem[{\citenamefont{Jacobs et~al.}(1995)\citenamefont{Jacobs, Rechenbach,
  and Zachwieja}}]{Jacobs:1995_JAC}
\bibinfo{author}{\bibfnamefont{H.}~\bibnamefont{Jacobs}},
  \bibinfo{author}{\bibfnamefont{D.}~\bibnamefont{Rechenbach}},
  \bibnamefont{and}
  \bibinfo{author}{\bibfnamefont{U.}~\bibnamefont{Zachwieja}},
  \bibinfo{journal}{J. Alloys Compd.} \textbf{\bibinfo{volume}{227}},
  \bibinfo{pages}{10} (\bibinfo{year}{1995}).

\bibitem[{Not({\natexlab{b}})}]{Note:errors}
\bibinfo{note}{For XAFS results, the error bars are the diagonal elements of
  the covariance matrix evaluated during the fit and rescaled by the square
  root of the reduced $\chi^2$. For the DFT method, they are the resulting
  variance of convergence tests.}

\bibitem[{\citenamefont{Swanson and Tatge}(1955)}]{Swanson:1955_NBoS}
\bibinfo{author}{\bibfnamefont{H.}~\bibnamefont{Swanson}} \bibnamefont{and}
  \bibinfo{author}{\bibfnamefont{E.}~\bibnamefont{Tatge}},
  \bibinfo{journal}{National Bureau of Standards}
  \textbf{\bibinfo{volume}{539}}, \bibinfo{pages}{1} (\bibinfo{year}{1955}).

\bibitem[{\citenamefont{Schulz and Thiemann}(1977)}]{Schulz:1977_SSC}
\bibinfo{author}{\bibfnamefont{H.}~\bibnamefont{Schulz}} \bibnamefont{and}
  \bibinfo{author}{\bibfnamefont{K.~H.} \bibnamefont{Thiemann}},
  \bibinfo{journal}{Solid State Commun.} \textbf{\bibinfo{volume}{23}},
  \bibinfo{pages}{815} (\bibinfo{year}{1977}).

\bibitem[{\citenamefont{Li et~al.}(2008)\citenamefont{Li, Simbrunner,
  Navarro-Quezada, Wegscheider, Quast, Litvinov, Gerthsen, and
  Bonanni}}]{Li:2008_JCG}
\bibinfo{author}{\bibfnamefont{T.}~\bibnamefont{Li}},
  \bibinfo{author}{\bibfnamefont{C.}~\bibnamefont{Simbrunner}},
  \bibinfo{author}{\bibfnamefont{A.}~\bibnamefont{Navarro-Quezada}},
  \bibinfo{author}{\bibfnamefont{M.}~\bibnamefont{Wegscheider}},
  \bibinfo{author}{\bibfnamefont{M.}~\bibnamefont{Quast}},
  \bibinfo{author}{\bibfnamefont{D.}~\bibnamefont{Litvinov}},
  \bibinfo{author}{\bibfnamefont{D.}~\bibnamefont{Gerthsen}}, \bibnamefont{and}
  \bibinfo{author}{\bibfnamefont{A.}~\bibnamefont{Bonanni}},
  \bibinfo{journal}{J. Cryst. Growth} \textbf{\bibinfo{volume}{310}},
  \bibinfo{pages}{3294 } (\bibinfo{year}{2008}).

\bibitem[{Not({\natexlab{c}})}]{Note:distortion}
\bibinfo{note}{The spectrum (5,300) presents a peak at k = 11 \AA$^{-1}$ that
  is unphysical. It is due to a distorsion by diffraction effects (as discussed
  in the experimental section) and a consequent wrong background removal.}

\bibitem[{\citenamefont{Giuli et~al.}(2002)\citenamefont{Giuli, Pratesi,
  Cipriani, and Paris}}]{Giuli:2002_GCA}
\bibinfo{author}{\bibfnamefont{G.}~\bibnamefont{Giuli}},
  \bibinfo{author}{\bibfnamefont{G.}~\bibnamefont{Pratesi}},
  \bibinfo{author}{\bibfnamefont{C.}~\bibnamefont{Cipriani}}, \bibnamefont{and}
  \bibinfo{author}{\bibfnamefont{E.}~\bibnamefont{Paris}},
  \bibinfo{journal}{Geochim. Cosmochim. Acta} \textbf{\bibinfo{volume}{66}},
  \bibinfo{pages}{4347} (\bibinfo{year}{2002}).

\bibitem[{\citenamefont{Malguth et~al.}(2006)\citenamefont{Malguth, Hoffmann,
  Gehlhoff, Gelhausen, Phillips, and Xu}}]{Malguth:2006_PRB}
\bibinfo{author}{\bibfnamefont{E.}~\bibnamefont{Malguth}},
  \bibinfo{author}{\bibfnamefont{A.}~\bibnamefont{Hoffmann}},
  \bibinfo{author}{\bibfnamefont{W.}~\bibnamefont{Gehlhoff}},
  \bibinfo{author}{\bibfnamefont{O.}~\bibnamefont{Gelhausen}},
  \bibinfo{author}{\bibfnamefont{M.~R.} \bibnamefont{Phillips}},
  \bibnamefont{and} \bibinfo{author}{\bibfnamefont{X.}~\bibnamefont{Xu}},
  \bibinfo{journal}{Phys. Rev. B} \textbf{\bibinfo{volume}{74}},
  \bibinfo{pages}{165202} (\bibinfo{year}{2006}).

\bibitem[{\citenamefont{Monemar}(1974)}]{Monemar:1974_PRB}
\bibinfo{author}{\bibfnamefont{B.}~\bibnamefont{Monemar}},
  \bibinfo{journal}{Phys. Rev. B} \textbf{\bibinfo{volume}{10}},
  \bibinfo{pages}{676} (\bibinfo{year}{1974}).

\bibitem[{\citenamefont{Malguth et~al.}(2008)\citenamefont{Malguth, Hoffmann,
  and Phillips}}]{Malguth:2008_PSSB}
\bibinfo{author}{\bibfnamefont{E.}~\bibnamefont{Malguth}},
  \bibinfo{author}{\bibfnamefont{A.}~\bibnamefont{Hoffmann}}, \bibnamefont{and}
  \bibinfo{author}{\bibfnamefont{M.~R.} \bibnamefont{Phillips}},
  \bibinfo{journal}{Phys. Status Solidi B} \textbf{\bibinfo{volume}{245}},
  \bibinfo{pages}{455} (\bibinfo{year}{2008}).

\bibitem[{\citenamefont{Dietl}(2008{\natexlab{a}})}]{Dietl:2008_PRB}
\bibinfo{author}{\bibfnamefont{T.}~\bibnamefont{Dietl}},
  \bibinfo{journal}{Phys. Rev. B} \textbf{\bibinfo{volume}{77}},
  \bibinfo{pages}{085208} (\bibinfo{year}{2008}{\natexlab{a}}).

\bibitem[{\citenamefont{Leineweber et~al.}(1999)\citenamefont{Leineweber,
  Jacobs, Huning, Lueken, Schilder, and Kockelmann}}]{Leineweber:1999_JAC}
\bibinfo{author}{\bibfnamefont{A.}~\bibnamefont{Leineweber}},
  \bibinfo{author}{\bibfnamefont{H.}~\bibnamefont{Jacobs}},
  \bibinfo{author}{\bibfnamefont{F.}~\bibnamefont{Huning}},
  \bibinfo{author}{\bibfnamefont{H.}~\bibnamefont{Lueken}},
  \bibinfo{author}{\bibfnamefont{H.}~\bibnamefont{Schilder}}, \bibnamefont{and}
  \bibinfo{author}{\bibfnamefont{W.}~\bibnamefont{Kockelmann}},
  \bibinfo{journal}{J. Alloys Compd.} \textbf{\bibinfo{volume}{288}},
  \bibinfo{pages}{79} (\bibinfo{year}{1999}).

\bibitem[{\citenamefont{Kumagai et~al.}(2006)\citenamefont{Kumagai, Satoh,
  Togashi, Murakami, Takemoto, Iihara, Yamaguchi, and
  Koukitu}}]{Kumagai:2006_JCG}
\bibinfo{author}{\bibfnamefont{Y.}~\bibnamefont{Kumagai}},
  \bibinfo{author}{\bibfnamefont{F.}~\bibnamefont{Satoh}},
  \bibinfo{author}{\bibfnamefont{R.}~\bibnamefont{Togashi}},
  \bibinfo{author}{\bibfnamefont{H.}~\bibnamefont{Murakami}},
  \bibinfo{author}{\bibfnamefont{K.}~\bibnamefont{Takemoto}},
  \bibinfo{author}{\bibfnamefont{J.}~\bibnamefont{Iihara}},
  \bibinfo{author}{\bibfnamefont{K.}~\bibnamefont{Yamaguchi}},
  \bibnamefont{and} \bibinfo{author}{\bibfnamefont{A.}~\bibnamefont{Koukitu}},
  \bibinfo{journal}{J. Cryst. Growth} \textbf{\bibinfo{volume}{296}},
  \bibinfo{pages}{11} (\bibinfo{year}{2006}).

\bibitem[{\citenamefont{Navarro-Quezada
  et~al.}(2008)\citenamefont{Navarro-Quezada, Li, Simbrunner, Kiecana,
  Hernandez-Sosa, Quast, Wegscheider, Sawicki, Dietl, and
  Bonanni}}]{Navarro-Quezada:2008_JCG}
\bibinfo{author}{\bibfnamefont{A.}~\bibnamefont{Navarro-Quezada}},
  \bibinfo{author}{\bibfnamefont{T.}~\bibnamefont{Li}},
  \bibinfo{author}{\bibfnamefont{C.}~\bibnamefont{Simbrunner}},
  \bibinfo{author}{\bibfnamefont{M.}~\bibnamefont{Kiecana}},
  \bibinfo{author}{\bibfnamefont{G.}~\bibnamefont{Hernandez-Sosa}},
  \bibinfo{author}{\bibfnamefont{M.}~\bibnamefont{Quast}},
  \bibinfo{author}{\bibfnamefont{M.}~\bibnamefont{Wegscheider}},
  \bibinfo{author}{\bibfnamefont{M.}~\bibnamefont{Sawicki}},
  \bibinfo{author}{\bibfnamefont{T.}~\bibnamefont{Dietl}}, \bibnamefont{and}
  \bibinfo{author}{\bibfnamefont{A.}~\bibnamefont{Bonanni}},
  \bibinfo{journal}{J. Cryst. Growth} \textbf{\bibinfo{volume}{310}},
  \bibinfo{pages}{1772} (\bibinfo{year}{2008}).

\bibitem[{\citenamefont{Dietl}(2008{\natexlab{b}})}]{Dietl:2008_JAP}
\bibinfo{author}{\bibfnamefont{T.}~\bibnamefont{Dietl}}, \bibinfo{journal}{J.
  Appl. Phys.} \textbf{\bibinfo{volume}{103}}, \bibinfo{pages}{07D111}
  (\bibinfo{year}{2008}{\natexlab{b}}).

\end{thebibliography}
\end{document}